\documentclass[
preprintnumbers,%
prc,%
10pt,%
final,%
notitlepage,%
oneside,%
twocolumn,%
nobibnotes,%
nofootinbib,
superscriptaddress,%
floatfix,%
showkeys,%
showpacs]%
{revtex4-1}
\usepackage{color} 
\usepackage{amsmath}
\usepackage{amssymb}
\usepackage{amsfonts}
\usepackage{amsbsy}
\usepackage{mathrsfs}
\usepackage{graphicx}
\usepackage{multirow}
\usepackage{lineno}

\def\lsim{\mathrel{\rlap{
\lower4pt\hbox{\hskip-3pt$\sim$}}
    \raise1pt\hbox{$<$}}}     
\def\gsim{\mathrel{\rlap{
\lower4pt\hbox{\hskip-3pt$\sim$}}
    \raise1pt\hbox{$>$}}}     

\preprint{draft \today}

\begin{document}


\title{Correlation femtoscopy study at energies available at the JINR Nuclotron-based Ion Collider fAcility and the BNL Relativistic Heavy Ion Collider within a viscous hydrodynamic plus cascade model}

\author{P. Batyuk}\thanks{e-mail: pavel.batyuk@jinr.ru}
\affiliation{Veksler and Baldin Laboratory of High Energy Physics, JINR Dubna, 141980 Dubna, Russia}
  
\author{Iu. Karpenko}
\affiliation{Bogolyubov Institute for Theoretical Physics, 03680 Kiev, Ukraine}
\affiliation{INFN - Sezione di Firenze, I-50019 Sesto Fiorentino (Firenze), Italy}

\author{R.Lednicky}
\affiliation{Veksler and Baldin Laboratory of High Energy Physics, JINR Dubna, 141980 Dubna, Russia}

\author{L.Malinina}
  \affiliation{Veksler and Baldin Laboratory of High Energy Physics, JINR Dubna, 141980 Dubna, Russia}
  \affiliation{M.~V.~Lomonosov Moscow State University, Moscow, Russia}
  \affiliation{D.~V.~Skobeltsyn Institute of Nuclear Physics, Moscow, Russia}
  
\author{K.Mikhaylov}
  \affiliation{Veksler and Baldin Laboratory of High Energy Physics, JINR Dubna, 141980 Dubna, Russia}
  \affiliation{Institute of Theoretical and Experimental Physics (ITEP), Moscow, Russia} 
  
\author{O. Rogachevsky}
\affiliation{Veksler and Baldin Laboratory of High Energy Physics, JINR Dubna, 141980 Dubna, Russia}

\author{D.Wielanek}
  \affiliation{Warsaw University of Technology, Faculty of Physics, Warsaw 00662, Poland}


\begin{abstract}
Correlation femtoscopy allows one to measure the space-time characteristics of particle production in relativistic heavy-ion collisions due to the effects of quantum statistics (QS) and final state interactions (FSI). 
The main features of the femtoscopy measurements at top RHIC and LHC energies are considered as a manifestation of strong collective flow and are well interpreted within hydrodynamic models employing equation of 
state (EoS) with a crossover type transition between Quark-Gluon Plasma (QGP) and hadron gas phases. The femtoscopy at lower energies was intensively studied at AGS and SPS accelerators and is being studied now in the Beam Energy Scan 
program (BES) at the BNL Relativistic Heavy Ion Collider in the context of exploration of the QCD phase diagram. In this article we present femtoscopic observables calculated for Au-Au collisions at 
$\sqrt{s_{NN}} = 7.7 - 62.4$~GeV in a viscous hydro + cascade model \texttt{vHLLE+UrQMD} and their dependence on the EoS of thermalized matter.

\end{abstract}

\maketitle

\section{Introduction}
One of the main motivations for heavy-ion collision programs is to study a new state of matter, the Quark-Gluon Plasma (QGP) which is defined as a deconfined state of quarks and gluons~\cite{QGP_a, QGP_b}. 
The systematics of transverse momentum spectra, elliptic and higher order flow coefficients measured in heavy-ion collisions at BNL Relativistic Heavy Ion Collider (RHIC) and CERN Large Hadron Collider (LHC) energies 
confirmed the presence of strong collective motion and hydrodynamic behavior of the system~\cite{RHIC_hydro_a, RHIC_hydro_b, RHIC_hydro_c, RHIC_hydro_d, ALICE_hydro_a, ALICE_hydro_b, ALICE_hydro_c, ALICE_hydro_d}.
While the hydrodynamic approach was successful in reproduction of elliptic flow measured at the top RHIC energies from the very beginning, it was unable to reproduce the femtoscopic correlation measurements. The problem was solved several years ago along with substantial improvements in hydrodynamic modeling~\cite{Sin09, AK08,Pratt09}. The improvements comprise a presence of pre-thermal transverse flow, an inclusion of shear viscous corrections to hydrodynamic evolution, an equation of state compatible with recent lattice QCD calculations, and a consistent treatment of hadronic stage (hadronic cascade phase).
As a result, existing state-of-the-art hydrodynamic models can reproduce, besides the transverse momentum distributions and elliptic flow coefficients, also the pion femtoscopic measurements~\cite{Werner:2010aa, Bosek_2011, Karpenko:2012yf}.

Recent lattice QCD calculations show that the transition from QGP to a hadron gas at high temperature and small $\mu_{B}$ is a crossover~\cite{Lat6_a, Lat6_b, Lat6_c, Lat6_d}, and there are no signs of critical behavior in the region of $\mu_{\rm B}/T<2$~\cite{Karsch:2015nqx}. This is supported by a recent analysis of combined full RHIC ($\sqrt{s_{NN}} = 200$~GeV) and LHC ($\sqrt{s_{NN}} = 2.76$~TeV) data in viscous hydrodynamic + cascade model, where an elaborate model-to-data comparison using Bayesian framework suggests that the speed of sound at all temperatures cannot fall below the hadron resonance gas value of $\sim 0.15$, and that the resulting posterior distribution over possible equations of state of matter is compatible with the lattice QCD results~\cite{Pratt:2015zsa}.
At the same time, there are predictions inspired by the lattice QCD calculations on a possible change of existing regime to a first-order phase transition occurring at lower energies and higher chemical potentials~\cite{ML10_16_a, ML10_16_b, ML10_16_c, ML10_16_d, ML10_16_e, ML10_16_f, ML10_16_g} and thus implying the existence of a critical point on the QCD phase diagram at a moderate value of chemical potential~\cite{ML17}. These considerations motivated the BES program allowing one to study different parts of the QCD phase diagram at existing accelerators like SPS and RHIC, and, in future, at NICA and FAIR facilities.

It has been shown many years ago~\cite{RG96_a, RG96_b, RG96_c} that the long duration of particle emission related to a first order phase transition
could reveal itself in the energy region of the onset of deconfinement as a strong increase of the Gaussian femtoscopic radius $R_{out}$, measured along the pair transverse momentum, compared with the nearly constant radius $R_{side}$, measured along the perpendicular direction in the transverse plane. As a result, one may expect a strong increase of the ratio $R_{out} / R_{side}$. In fact, a first order phase transition leads to a stalling of the mean expansion speed and a longer emission duration $\Delta\tau$ manifested as an increase of the radius measured along the beam direction $R_{long}$ and of the ratio of transverse femtoscopy radii $R_{out} / R_{side}$, respectively.
A large data set of correlation functions of identical charged pions has been recently obtained by the STAR Collaboration within the RHIC BES at $\sqrt{s_{NN}} = $ 7.7, 11.5, 19.6, 27, 39, 62.4~GeV. The study of $R_{out} / R_{side}$ and $R_{out}^{2} - R_{side}^{2}$ behavior as a function of $\sqrt{s_{NN}}$ indicates a wide maximum near $\sqrt{s_{NN}} \sim$~20~GeV, which is reported also by the PHENIX Collaboration~\cite{PHENIX_HBT}. 
Could this wide maximum be related to the expected change of the type of phase transition?

To answer this and other questions, we study the sensitivity of Bose-Einstein correlations of identical pions to the EoS using a hybrid \texttt{vHLLE+UrQMD} model~\cite{vHLLE_UrQMD_Yura1}. The model combines the UrQMD approach~\cite{UrQMD_a, UrQMD_b} for the early and late stages of the evolution with numerical $(3 + 1)$-dimensional viscous hydrodynamical solution~\cite{vHLLE} for the hot and dense expanding matter. A hydrodynamic approach has an essential advantage for the present analysis, since it allows one to simulate different scenarios of hadron / quark-gluon transition by changing the EoS and other transport coefficients input.

The paper is organized as follows: the details of the model are discussed in Section~\ref{sec:sec1}; in Section~\ref{sec:sec2} the femtoscopy formalism is described; in Section~\ref{sec:sec3} the results are presented and discussed; Section~\ref{sec:sec4} is dedicated to conclusions.

\section{vHLLE+UrQMD model}
\label{sec:sec1}
The use of multi-component dynamical models for the description of dynamics of relativistic heavy-ion collisions at RHIC and higher energies is essential because hydrodynamics alone cannot describe the entire reaction. At the early stage of collision, thermalization of out-of-equilibrium quark-gluon system is assumed to occur, which allows one to describe the subsequent complex multi-particle dynamics using a relatively simple formalism of relativistic hydrodynamics.
This requires an approach to calculate initial conditions for the hydrodynamic evolution\footnote{Note that for lower collision energies there exist one-, two-, and three-fluid hydrodynamical models (e.g.~\cite{Ivanov}), which apply hydrodynamical description already for incoming nuclei.}. 
As the matter expands, the characteristic mean free path of its constituents (quarks and gluons transforming into hadrons) becomes comparable to the system size. The interactions become less frequent, but they do not cease instantaneously, a process which can be simulated by switching from the hydrodynamical evolution to a hadronic cascade, usually with the help of Cooper-Frye formula~\cite{CooperFrye}.

For the early stage of collision different approaches (or models) have been used in literature, such as \texttt{(MC-)KLN}, \texttt{IP-Glasma}, \texttt{EPOS}, \texttt{HIJING}, Glauber model etc. Those approaches have been developed for full RHIC or LHC energies, and lose their applicability as the collision energy decreases. As for the Glauber model, it can estimate initial energy density profiles in transverse direction only. Therefore we choose to use UrQMD to simulate the initial stage of the collision. We enforce a transition to hydrodynamical description at a hyper-surface of constant longitudinal proper time $\tau_0=\sqrt{t^2-z^2}$. The minimal value of the starting time $\tau_0$ is taken to be equal to the average time for the two colliding nuclei to completely pass through each other:

\begin{equation}
\tau_0=2R/\sqrt{(\sqrt{s_{\rm NN}}/2m_N)^2-1}, \label{eqTau0}
\end{equation}
where R is average radius of the nucleus and $m_N$ is nucleon mass.

At $\tau=\tau_0$ energy, momentum and baryon/electric charges of hadrons are distributed to fluid cells $ijk$ around each hadron's position according to Gaussian profiles:
\begin{align} \label{Gauss1}
\Delta P^\alpha_{ijk} & = P^\alpha \cdot C\cdot\exp\left(-\frac{\Delta x_i^2+\Delta y_j^2}{R_\perp^2}-\frac{\Delta\eta_k^2}{R_\eta^2}\gamma_\eta^2 \tau_0^2\right) \\
\Delta N^0_{ijk}&=N^0 \cdot C\cdot\exp\left(-\frac{\Delta x_i^2+\Delta y_j^2}{R_\perp^2}-\frac{\Delta\eta_k^2}{R_\eta^2}\gamma_\eta^2 \tau_0^2\right), \label{Gauss2}
\end{align}
where $P^\alpha$ and $N^0$ are 4-momentum and charge of a hadron, $\{\Delta x_i, \Delta y_j, \Delta \eta_k\}$ are the distances between hadron's position and center of a hydro cell $ijk$ in each direction, $\gamma_\eta={\rm cosh}(y_p-\eta)$ is the longitudinal Lorentz factor of the hadron as seen in a frame moving with the rapidity $\eta$, and $C$ is a normalization constant. The normalization constant $C$ is calculated so that the discrete sum of energy depositions to the hydrodynamic cells equals to the energy of the hadron. The width parameters $R_\perp$ and $R_\eta$ control granularity of the produced initial state.

For all collision energies in consideration, the resulting initial energy density is large enough for the dense parts of the system to reside in the QGP phase.

\begin{figure}
\includegraphics[width=0.5\textwidth]{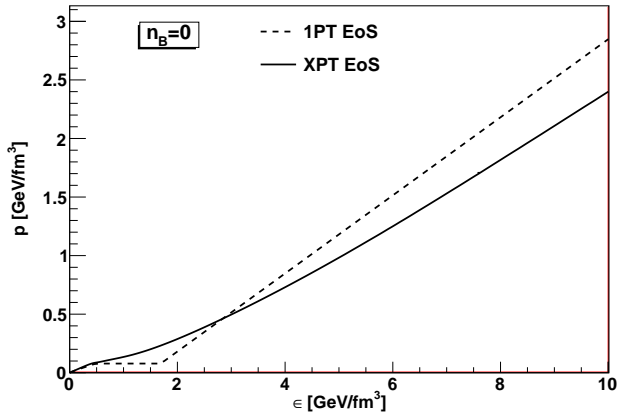}
\caption{Thermodynamic pressure as a function of energy density, evaluated at zero baryon density from the equations of state used in the hydrodynamic stage: chiral model EoS with crossover transition (XPT) and bag model EoS with first order phase transition (1PT).}\label{fig-eos}
\end{figure}

The following 3-dimensional viscous hydrodynamic expansion is simulated with the \texttt{vHLLE} code \cite{vHLLE}. Another input to the hydrodynamic part is the EoS, for which we use chiral model EoS~\cite{SchrammSteinheimer} or bag model EoS~\cite{KolbSollfrankHeinz}. Whereas the present version of the chiral model EoS has a crossover type transition between QGP and hadronic phases for all baryon densities, the bag model EoS has a first order phase transition between the phases also for all baryon densities. Therefore below we dub chiral model EoS as ``XPT EoS'', and bag model EoS as ``1PT EoS''. For both EoS there are publicly available tables, computed in full physically allowed $T-\mu_B$ region, which makes them particularly useful for hydrodynamic computations with fluctuating initial conditions. Pressure as a function of energy density from both EoS is demonstrated on Fig.~\ref{fig-eos}.

Fluid to particle transition, or particlization, is set to happen at a hypersurface of constant (hydrodynamic) energy density $\epsilon_{\rm sw}=0.5$~GeV/fm$^3$, when the hydrodynamic EoS corresponds to hadronic phase. The particlization hypersurface is reconstructed with the CORNELIUS subroutine~\cite{Huovinen:2012is}. At this hypersurface, individual hadrons are sampled using the Cooper-Frye formula including shear viscous corrections to the distribution functions. The hadronic rescatterings and decays are treated with the \texttt{UrQMD} cascade.

The initial state parameters $R_\perp$, $R_\eta$, hydrodynamic starting time $\tau_0$ and shear viscosity over entropy ratio $\eta/s$ in fluid phase are tuned for different collision energies in order to approach basic experimental observables in the RHIC Beam Energy Scan region: (pseudo)rapidity distributions, transverse momentum spectra and elliptic flow coefficient \cite{vHLLE_UrQMD_Yura1}. The resulting values of the parameters are presented in Table~\ref{tbParams}. The tuning has been made with the XPT (chiral model) EoS, and in present work we use the same set of parameter values (i.e.\ do not re-tune them) for the simulations with the 1PT (bag model) EoS.

\begin{table}
\begin{tabular}{|l|l|l|l|l|}
\hline
 $\sqrt{s_{\rm NN}}$~[GeV] & $\tau_0$~[fm/c] & $R_\perp$~[fm] & $R_\eta$~[fm] & $\eta/s$ \\ \hline
     7.7          &      3.2        &     1.4        &     0.5    &    0.2   \\ \hline
     8.8 (SPS)    &      2.83       &     1.4        &     0.5    &    0.2   \\ \hline
     11.5         &      2.1        &     1.4        &     0.5    &    0.2   \\ \hline
     17.3 (SPS)   &      1.42       &     1.4        &     0.5    &    0.15  \\ \hline
     19.6         &      1.22       &     1.4        &     0.5    &    0.15  \\ \hline
     27           &      1.0        &     1.2        &     0.5    &    0.12  \\ \hline
     39           &      0.9*        &     1.0        &     0.7    &    0.08  \\ \hline
     62.4         &      0.7*        &     1.0        &     0.7    &    0.08  \\ \hline
     200          &      0.4*        &     1.0        &     1.0    &    0.08  \\ \hline
 \end{tabular}
\caption{Values of hydrodynamic starting time $\tau_0$, initial state granularity $R_\perp$, $R_\eta$ and shear viscosity over entropy ratio $\eta/s$ adjusted for different collision energies in order to reproduce basic observables in the RHIC BES region. An asterisk marks the values of $\tau_0$ which are adjusted instead of being set directly from Eq.~\ref{eqTau0}.}\label{tbParams}
\end{table}

\section{Femtoscopy formalism}
\label{sec:sec2}
Since the first demonstration of the sensitivity of the Bose-Einstein correlations to the spatial scale of the emitting source done almost 60 years ago by G. Goldhaber, S. Goldhaber, W. Lee and A. Pais~\cite{GGLP}, the momentum correlation technique was successfully developed and is known presently as a ``correlation femtoscopy''.
It was successfully applied to the measurement of the space-time characteristics of particle production processes in high energy collisions, especially in heavy-ion collisions~\cite{pod89_a, pod89_b, led04, lis05_a, lis05_b}. 
Femtoscopy correlations are studied by means of a two-particle correlation function. 
In a production process of a small enough phase space density, the
correlations of two particles emitted with a small momentum 
$k^*= |{\bf k}^*|$ 
in the pair rest frame (PRF)\footnote{Calculations made in PRF are denoted by asterisk.} are dominated by the effects of their
mutual final state interaction (FSI) and quantum statistics (QS),
depending on the PRF temporal ($t^* = t_1^* - t_2^*$) and spatial (${\bf r}^* = {\bf r}_1^* - {\bf r}_2^*$) separation of
particle emission points. Usually, one can neglect the temporal separation~\cite{Lednicky:1981su, Lednicky:2005tb} and in such equal-time approximation describe these effects by properly symmetrized wave functions at a given total pair spin $\cal{S}$,
$\left[\psi_{-{\bf k}^*}^{\cal{S},\alpha^{\prime}\alpha}({\bf r}^*)\right]^*$,
representing solutions of the scattering problem viewed in the opposite
time direction. So the complex conjugate, negative sign of the vector
${\bf k}^* = {\bf p}_1^* = -{\bf p}_2^*$ and the detected channel $\alpha$ being the entrance one.
Since the FSI factorization requires the FSI duration much larger than the
particle production time, the relative momentum should be small also in
the intermediate channels $\alpha^{\prime}$, so that one may consider the
particles in these channels belonging to the same isospin multiplets as
those in the detected channel $\alpha$.
Particularly, for identical particles, the multi-channel problem reduces
to the single elastic transition $\alpha \to \alpha$ only. Assuming
further sufficiently smooth behavior of single-particle spectra in a
narrow correlation region (smoothness assumption)~\cite{pod89_a,pod89_b}, one can neglect
the space-time coherence and write the correlation function at a given
${\bf k}^*$ and pair three-momentum ${\bf P}$ as
\begin{equation}
C({\bf k}^*,{\bf P})= \int {\rm d}^3{\bf r}^* S^\alpha({\bf r}^*,{\bf P})
                                 \overline{\left| \psi_{-{\bf
k}^*}^{\cal{S},\alpha^{\prime}\alpha}({\bf r}^*)
\right|^2},
\label{Koonin_Pratt}
\end{equation}
where the overline describes the averaging over the total pair spin $\cal{S}$
and summing over the intermediate channels $\alpha^{\prime}$. It is implied that particles are produced in a complex process with equilibrated spin and
isospin projections and so the separation distribution (source function)
$S^\alpha({\bf r}^*,{\bf P})$
is independent of $\cal{S}$ and $\alpha^{\prime}$.

Experimentally, a two-particle correlation function is defined as a ratio $C({\bf q}) = A({\bf q}) / B({\bf q})$.
$A({\bf q})$ is a measured distribution of the difference ${\bf q} = {\bf p}_1 - {\bf p}_2$, where ${\bf p}_1$ and ${\bf p}_2$ are three-momenta of two considered particles taken from the same event, while $B({\bf q})$ is a reference distribution of pairs of particles taken from different events.
The momentum difference is usually calculated in the longitudinally co-moving system (LCMS), where the longitudinal pair momentum vanishes. The vector ${\bf q}$ is usually expressed in terms of $q_{\rm out}$ , $q_{\rm side}$ , $q_{\rm long}$, where the ``long'' axis is directed along the beam, ``out'' - along the pair transverse momentum and ``side'' -  perpendicular to the latter one
in the transverse plane. 

To perform a quantitative analysis of femtoscopic correlations, an analytical form of $S$ is often used so that the result of the integration procedure 
in Eq.~(\ref{Koonin_Pratt}) can be compared with a correlation function 
$C$ obtained from an experiment. The source function is usually considered independent of the relative momentum ${\bf q}$ and its Gaussian shape is assumed:
\begin{equation}
S(\bf{r}) \sim  exp \left(-\frac{{r^*}_{out}^{2}}{4{R^*}_{out}^{2}}-\frac{{r^*}_{side}^{2}}{4{R^*}_{side}^{2}} 
-\frac{{r^*}_{long}^{2}}{4{R^*}_{long}^{2}} \right).
\label{SF_Gaussian}
\end{equation}
The widths in three directions (out, side and long) are called ``Gaussian femtoscopy radii''. 
In LCMS $q_{\rm out}=\gamma_{t} (q_{out}^{*}+\beta_t (m_1^2-m_{2}^{2})/m_{12})$,
               $q_{side}=q_{side}^*$, $q_{long}=q_{long}^*$, where
$\gamma_t$ and $\beta_t$ are the LCMS Lorentz factor and velocity of the
pair. Since the space-time separation in PRF and LCMS is related
by the Lorentz boost in the out direction: $r_{\rm out}^{*}=\gamma_t(r_{\rm out}-\beta_{t} t)$, $r_{\rm side}^{*}=r_{\rm side}$, $r_{\rm long}^{*}=r_{\rm long}$, the PRF and LCMS Gaussian radii coincide except for $R_{\rm out}^{*}=\gamma_{t} R_{\rm out}$.
In present paper we consider the correlation function of
two identical pions neglecting their FSI, so that
\begin{eqnarray}
\label{weight}
& \left|\psi_{-{\bf k}^*}({\bf r}^*)\right|^2=
            \left|[\exp(-i{\bf k}^*{\bf r}^*)+
             \exp(i{\bf k}^*{\bf r}^*)]/\sqrt{2}\right|^2 = \\ \nonumber
&            = 1 +\cos(2{\bf k}^*{\bf r}^*),    
\end{eqnarray}
and Eqs.~(\ref{Koonin_Pratt}, \ref{SF_Gaussian}, \ref{weight}) yield the 3-dim Gaussian form of the
correlation function. This form is usually used to fit the LCMS
Gaussian radii according to:
\begin{equation}
{C}({\bf q})= N \left(1+\lambda\exp(-R_{\rm out}^2q_{\rm out}^2-R_{\rm side}^2q_{\rm side}^2-R_{\rm long}^2q_{\rm long}^2)\right).
\label{eq:osl}
\end{equation}
where $N$ is the normalization factor and $\lambda$ is the
correlation strength parameter, which can differ from unity due to the
contribution of long-lived emitters and a non-Gaussian shape of the
correlation function; $R_{out}$, $R_{side}$, $R_{long}$ are the Gaussian femtoscopy radii in the LCMS frame. Eq.~(\ref{eq:osl}) assumes azimuthal symmetry of the production process, which forbids the presence of the cross-terms except for $q_{out} q_{long}$. We neglect the latter assuming further the invariance under longitudinal boosts. Generally, in case of a correlation analysis with respect to the reaction plane, all three cross-terms $q_{i}q_{j}$ contribute.

The described fitting procedure allows one to compare extracted femtoscopy radii from the model with existing experimental data. This can be considered as a standard approach. A disadvantage of this approach is due to the fact that the Gaussian parametrization can suppress important information that could be derived from the long non-Gaussian tails of source functions. 
The PHENIX and STAR collaborations have recently started to apply a
new ``imaging technique'' in order to extract directly the source function~\cite{SF}. In contrast to the standard approach, the source imaging allows one to extract a real non-Gaussian source function, being in this sense a model-independent one. 

Here we will use the \texttt{vHLLE+UrQMD} model to study the effect of a non-Gaussian shape of the source function and its dependence on the nature of the phase transition. This model provides the information on particle four-momenta and four-coordinates of the emission points allowing one to calculate the correlation function with the help of the weight procedure. For non-interacting identical pions, the weight is given in Eq.~(\ref{weight}).

\section{Results and Discussion}
\label{sec:sec3}
As mentioned above, the parameters of the model were adjusted to approach experimental data for (pseudo-)rapidity distributions, transverse momentum spectra and elliptic flow coefficients with the XPT EoS, corresponding to a crossover type transition \cite{vHLLE_UrQMD_Yura1}. Then, switching the EoS from XPT to 1PT leads to only small differences in multiplicities and rapidity distributions of the produced hadrons in the model. At the same time, hydrodynamic evolution with the 1PT EoS leads to somewhat decreased mean $p_\perp$, elliptic and triangular flow coefficients~\cite{Karpenko:2016red}. Those trends are explained by a less violent transverse expansion with the 1PT EoS.

In this section, we study the space-time characteristics of the hadron emission in the model and present the results for the Bose-Einstein correlations of identical pions, obtained with the two aforementioned EoS's in a wide collision energy range $\sqrt{s_{NN}}$ = 7.7 - 62.4~GeV covered by the BES program at RHIC.

\subsection{Pion emission time distributions}
\begin{figure}[!th]
  \begin{center}
    \includegraphics[width=.49\textwidth]{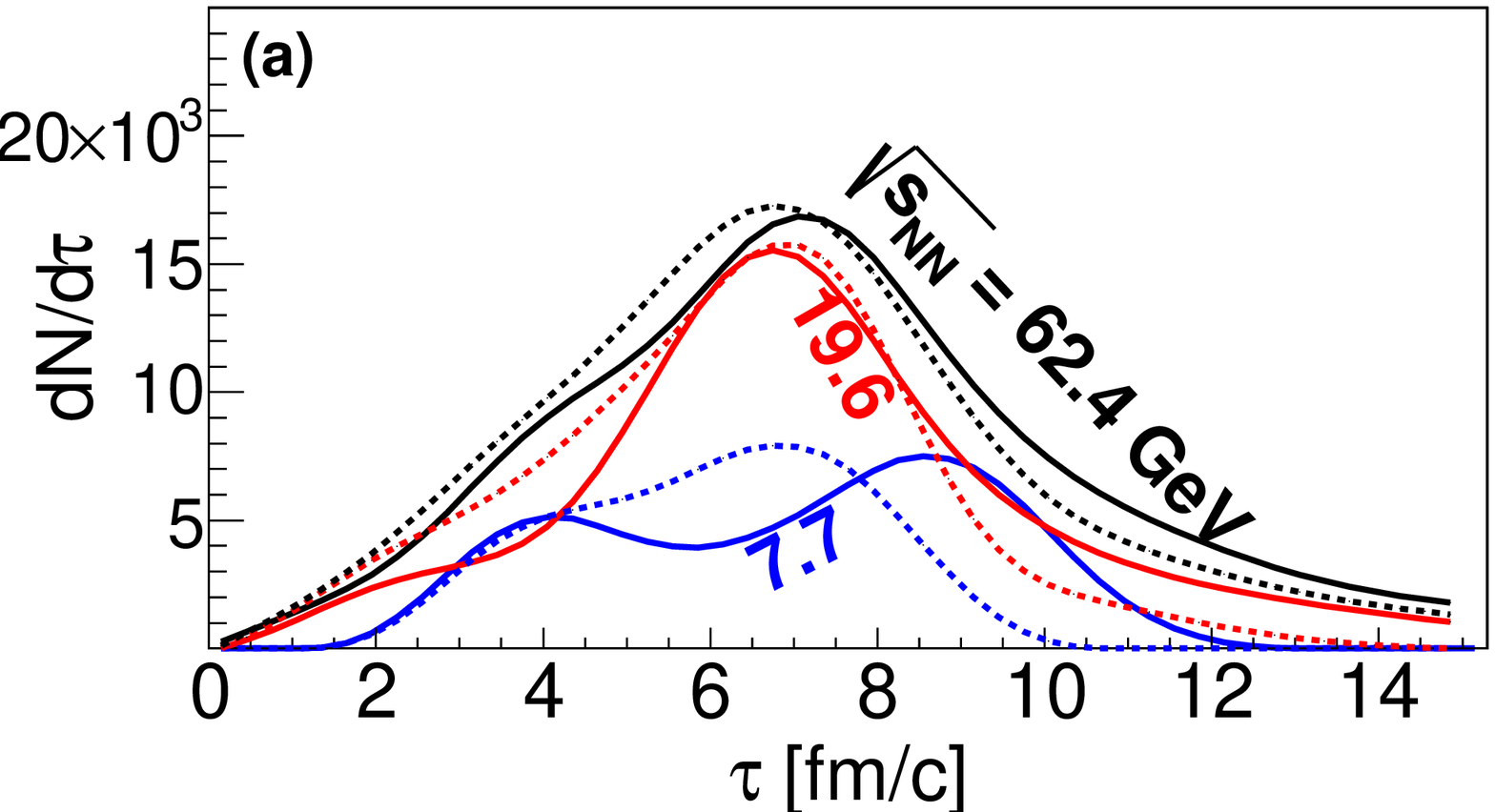}
    \includegraphics[width=.49\textwidth]{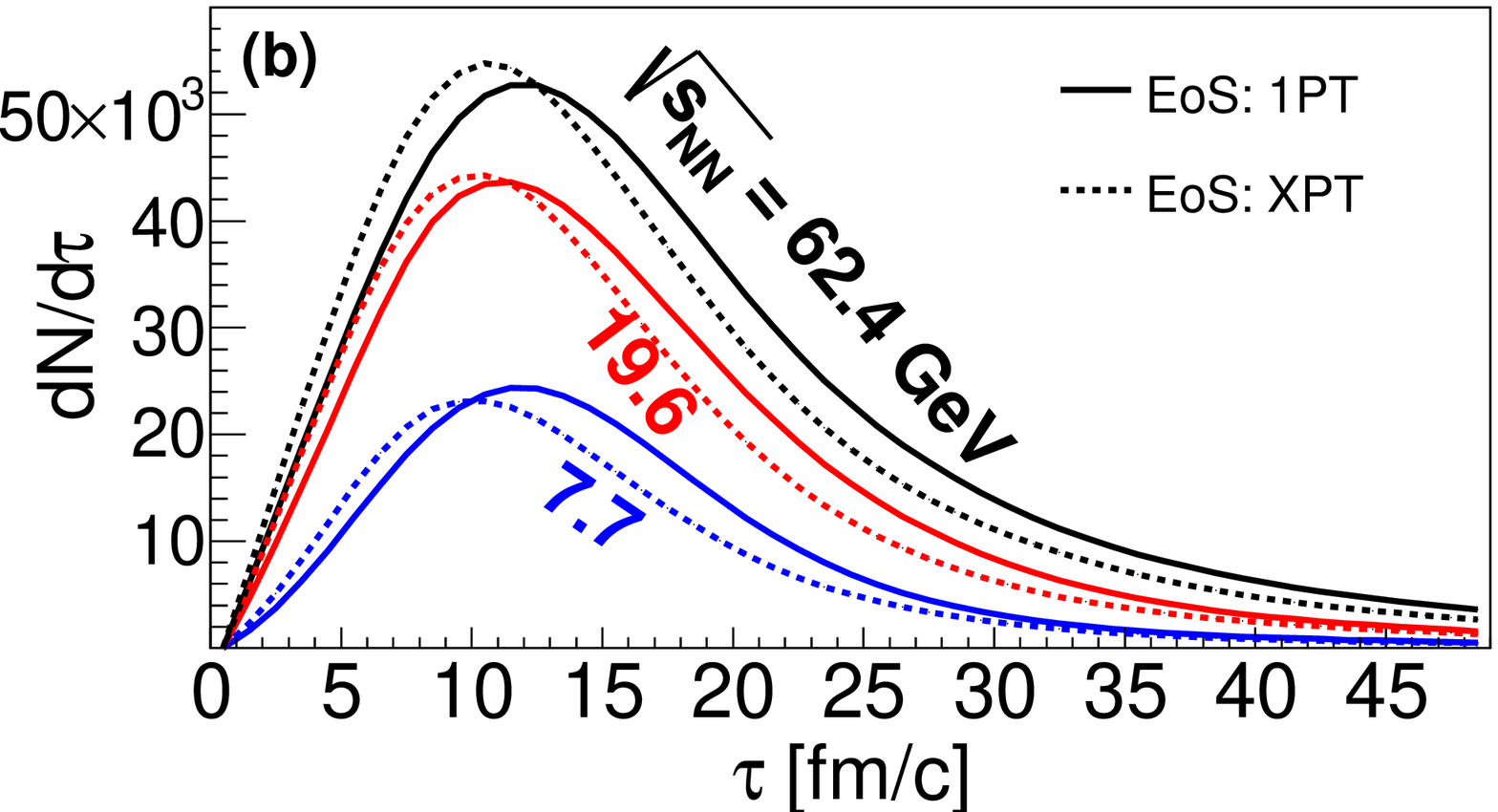}
    \caption{Pion emission times at the particlization surface (a) and the last interactions (b) in the center-of-mass system of colliding gold nuclei at different values of $\sqrt{s_{NN}}$.}  
    \label{fig:time_vHLLE}
  \end{center}
\end{figure}

In the model one has access to the space-time points of particle production 
in last collisions and resonance decays, in addition to their momenta.
In Fig.~\ref{fig:time_vHLLE} we visualize the averaged time distributions of last interaction points of pions from the model simulations at the lowest, middle and highest collision energies ($\sqrt{s_{NN}} = 7.7$, $19.6$ and $62.4$~GeV, respectively) using the 1PT and XPT EoS's.

A detailed information on the pion emission times as a function of $\sqrt{s_{NN}}$ for all simulated collision energies and EoS's is given in Table~\ref{tab:tau}. The time distributions of midrapidity pions have been obtained at the particlization surface (points of their creation) and at the points of last interactions.

\begin{table}[!h]
\caption{Extracted average pion emission times $\bar{t}$ as a function of $\sqrt{s_{NN}}$ in the center-of-mass system of colliding gold nuclei depending on the EoS used.}
\begin{center}
\begin{tabular}{|c|c|c|c|c|c|}\hline
$\sqrt{s_{NN}}$ & \multirow{2}{*}{EoS} & \multicolumn{2}{c|}{particlization surface} & \multicolumn{2}{c|}{last interactions}\\
\cline{3-6} 
$[GeV]$& & $\bar{t}$ [fm/c] &  RMS [fm/c] & $\bar{t}$ [fm/c] & RMS [fm/c]\\
\hline
  \multirow{2}{*}{7.7}     & 1PT     &  7.24 & 2.84 & 13.15    & 6.56    \\
  & XPT   &  6.16 & 2.01 & 11.61    & 6.26 \\
\hline
  \multirow{2}{*}{11.5}       & 1PT     &  7.33 & 2.31 & 13.09    & 6.92    \\
  & XPT  &  6.36 & 1.91  & 11.57   & 6.41   \\
\hline
  \multirow{2}{*}{19.6}       & 1PT     & 6.88 & 2.16 & 13.18     &  7.56  \\
 & XPT  & 6.41 & 2.15 & 11.93    &    6.93  \\
\hline
  \multirow{2}{*}{27}       & 1PT     &   6.85 & 2.37 & 13.38   &  8.07  \\
   & XPT  &    6.40 & 2.39 & 12.62 &    7.57   \\
 \hline
 \multirow{2}{*}{39}        & 1PT     &   7.17 & 2.75 & 13.98   &  8.30   \\
   & XPT  &    6.64 & 2.58 & 13.05 &    7.85   \\
\hline
 \multirow{2}{*}{62.4}        & 1PT     &   7.00 & 2.82 & 14.11   &  8.50  \\
   & XPT  &    6.60 & 2.63 & 12.72 &    7.81 \\     
\hline
\end{tabular}
\end{center}
\label{tab:tau}
\end{table}

From the Table~\ref{tab:tau} one can conclude an apparently weak dependence of the average pion creation time $\bar t$ at the particlization surface on collision energy.  It is an interplay of longer pre-thermal and shorter hydrodynamic stage at lower collision energies: at $\sqrt{s_{NN}} < 39$~GeV, the hydro stage starts at $\tau_0 = 2R/(\gamma v_z)$ when the two colliding nuclei have completely passed through each other, and the value of $\tau_0$ is as large as 3.2~fm/c at $\sqrt{s_{NN}} = 7.7$~GeV. On the other hand, the duration of hydro stage becomes shorter as a collision energy decreases because of lower initial energy density at the hydro starting time.
For the average pion creation times, the differences between 1PT EoS and XPT EoS  are largest at the lowest collision energy in consideration. In addition, the 1PT EoS leads to larger root-mean-square (RMS) values of the time distributions, and the difference is again largest at the lowest collision energy.
Because of re-scatterings and, more importantly, resonance decays in the final stage of hadronic cascade, the points of last interactions correspond to larger values of $\bar t$, which also depend weakly on collision energy. The cascade somewhat smears the relative difference between the 1PT and XPT scenarios, both for $\bar t$ and $RMS$.

\subsection{Three-dimensional correlation radii in LCMS.}
\begin{figure}[!th]
  \begin{center}
    \includegraphics[width=0.5\textwidth]{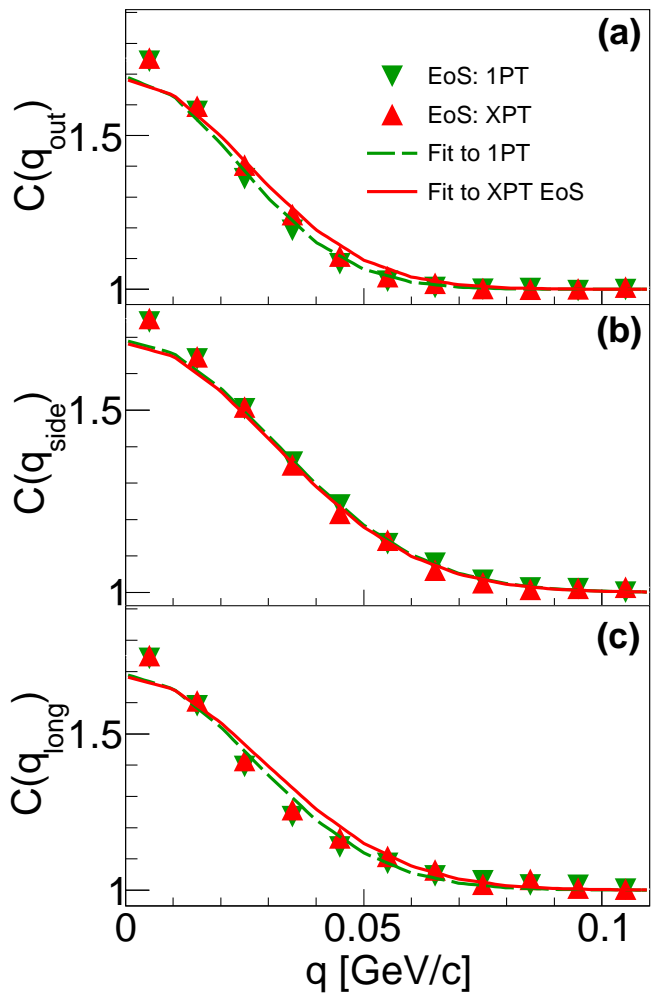}
    \caption{One-dimensional projections of three-dimensional correlation function (see Eq.~(\ref{eq:osl})) of non-interacting identical pion pairs onto ``out'' (a), ``side'' (b) and ``long'' (c) directions.  While projecting onto a direction, other two directions are required to be within the range of (-0.03, 0.03) GeV/c. A fit with the Gaussian function is presented by dashed and solid lines for the 1PT and XPT scenarios, respectively.} 
 \label{fig:CF}
  \end{center}
\end{figure}

An example of one-dimensional projections of three-dimensional correlation function obtained with the \texttt{vHLLE+UrQMD} model using the 1PT EoS and XPT EoS is shown in Fig.~\ref{fig:CF}. The analysis involved the  simulations performed for gold-gold collisions at $\sqrt{s_{NN}}= 7.7$ GeV with applied cuts on event centrality of $0-5\%$ and the pair transverse momentum $k_{T}$. The second one pertains to the range of (0.15 - 0.25) GeV/c. 

\begin{figure*}[!th]
  \begin{center}
    \includegraphics[width=0.80\textwidth]{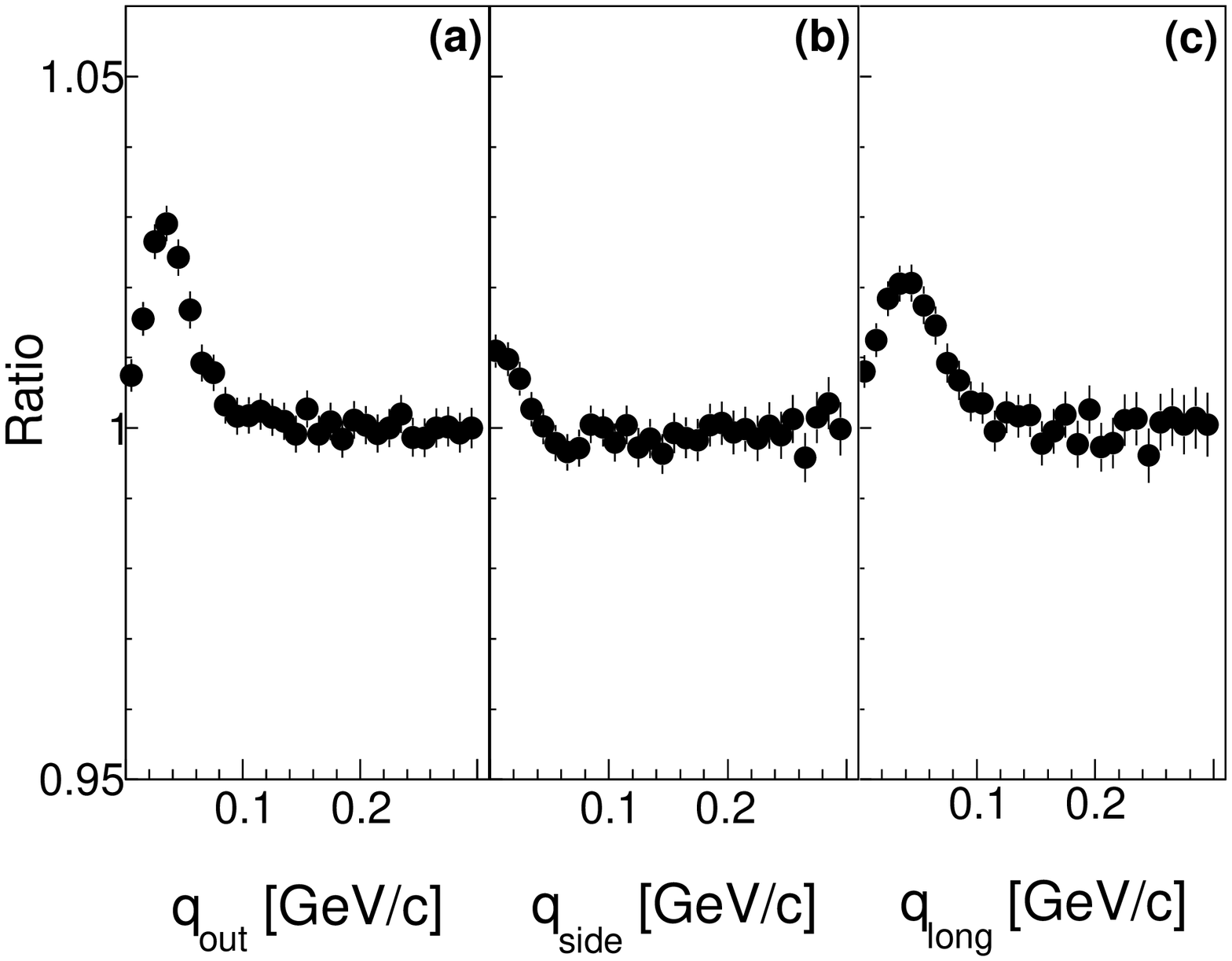}
    \caption{Ratios of one-dimensional projections of three-dimensional correlation functions for the two EoS. For each direction the corresponding ratio is calculated as follows: $C(q_{i}) (XPT) / C(q_{i}) (1PT)$, where $i$ denotes ``out'' (a), ``side'' (b) and ``long'' (c) directions, $1PT$ and $XPT$ denote a type of the used EoS.}
    \label{fig:CF_ratios}
  \end{center}
\end{figure*}

One can see that the pion correlation functions at small $q_{i}$ (i = ``out'', ``side'', ``long'') are not well described by the Gaussian function in Eq.~(\ref{eq:osl}). The observed difference between correlation functions calculated with the 1PT and XPT EoS's  is noticeable in the ``out'' and ``long''
directions. In Fig.~\ref{fig:CF_ratios} this fact is demonstrated by the ratios of individual projections. The ratios in the ``out'' and ``long'' directions reach values up to 1.03 at small $q_{out}$ and $q_{long}$. A percent deviation from unity at small $q_{side}$ values
appears due to the finite cuts on $q_{out}$ and $q_{long}$.

\begin{figure*}[!th]
  \begin{center}
    \includegraphics[width=1.\textwidth]{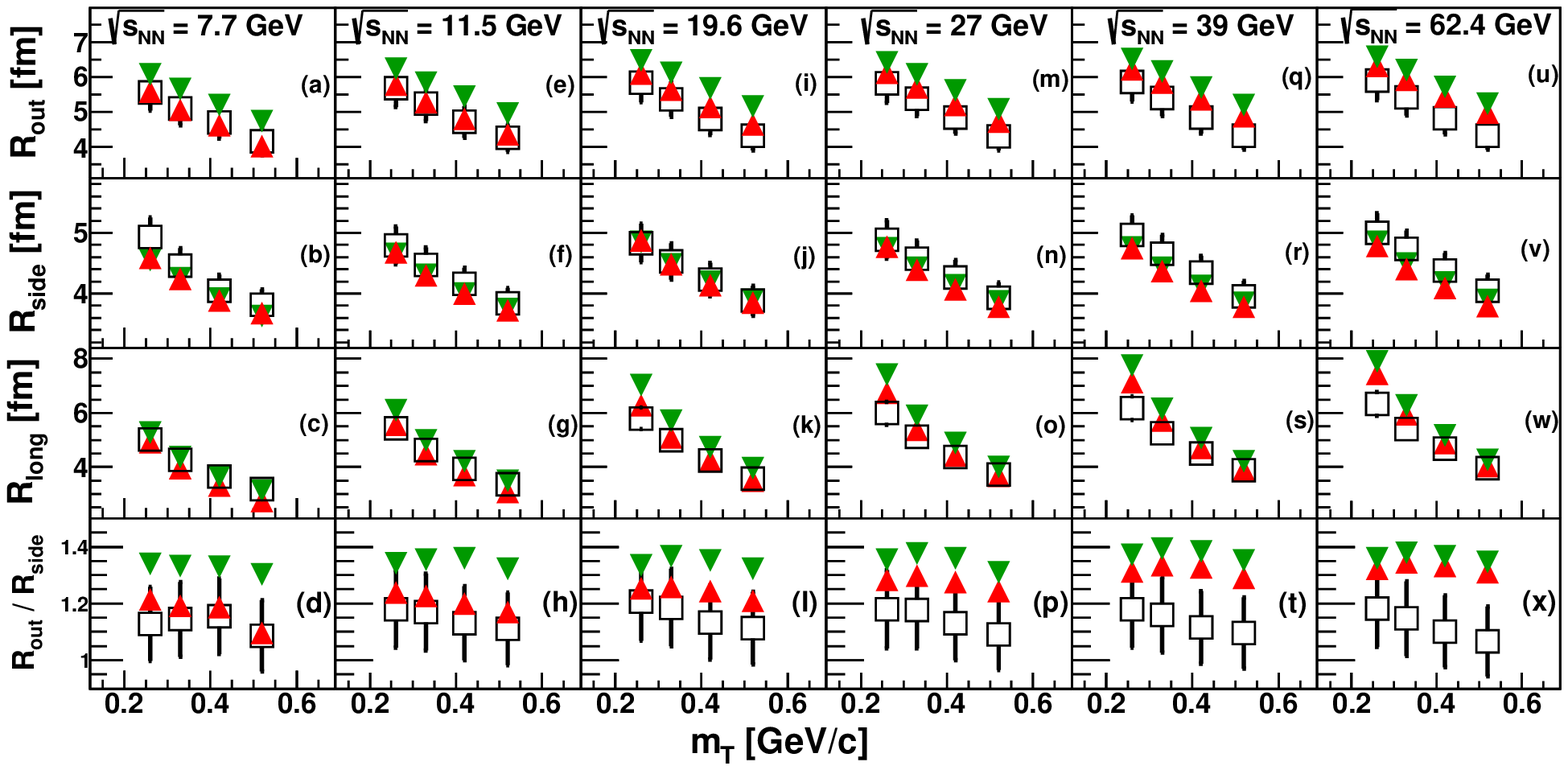}
    \caption{Comparison of the model three-dimensional LCMS femtoscopy radii fitted according to Eq.~(\ref{eq:osl}) with those measured by the STAR collaboration at $\sqrt{s_{NN}}$ = 7.7~((a) - (d)), 11.5~((e) - (h)), 19.6~((i) - (l)), 27~((m) - (p)), 39~((q) - (t)), 62.4~((u) - (x))~GeV. Open squares represents the STAR data. Triangles correspond to different types of EoS like they do in Fig.~\ref{fig:CF}.
    }\label{fig:R3Dall}
  \end{center}
\end{figure*}

In Fig.~\ref{fig:R3Dall} we present the $m_{T}$-dependence of the three-dimensional femtoscopy LCMS radii calculated at $\sqrt{s_{NN}}$ = 7.7, 11.5, 19.6, 27, 39, 62.4~GeV using the 1PT and XPT EoS's, and a comparison of the obtained results with those ones obtained by the STAR collaboration~\cite{STAR_HBT}.

One can see that the model reasonably describes the $m_{T}$-dependence of radii for all beam energies with both EoS's.  
 As for the radii, they show different trends in the ``out'', ``side'' and ``long'' directions. Whereas the $R_{side}$ using both EoS's practically coincide, the $R_{\rm out}$ with the 1PT EoS is generally larger (however, not more than 0.5~fm at any collision energy) than for the XPT EoS. This also leads to larger values of the 
 $R_{out} / R_{side}$ ratio using the 1PT EoS. 
 The difference comes from a weaker transverse flow developed in the fluid phase with the 1PT EoS as compared with the XPT EoS\footnote{A similar influence of the transverse flow on $R_{out}$ and $R_{side}$ has been observed for the RHIC and LHC energies~\cite{Karpenko:2012yf}.}. A longer lifetime of the fluid phase in the 1PT scenario also results in a larger values of $R_{long}$ as compared with the XPT scenario. Whereas one could expect that at lower collision energies in the 1PT EoS a larger fraction of the fluid phase evolution occurs in the mixed phase with zero speed of sound leading to  an increase of evolution time and $R_{long}$, we did not observe such a trend in the model. The reason is that at lower collision energies in the model a sizable amount of radial flow is developed at pre-hydro stage. At the same time, the $R_{out} / R_{side}$ ratio at lowest collision energies shows a clear EoS dependence.

\begin{figure}[!th]
  \begin{center}
    \includegraphics[width=0.5\textwidth]{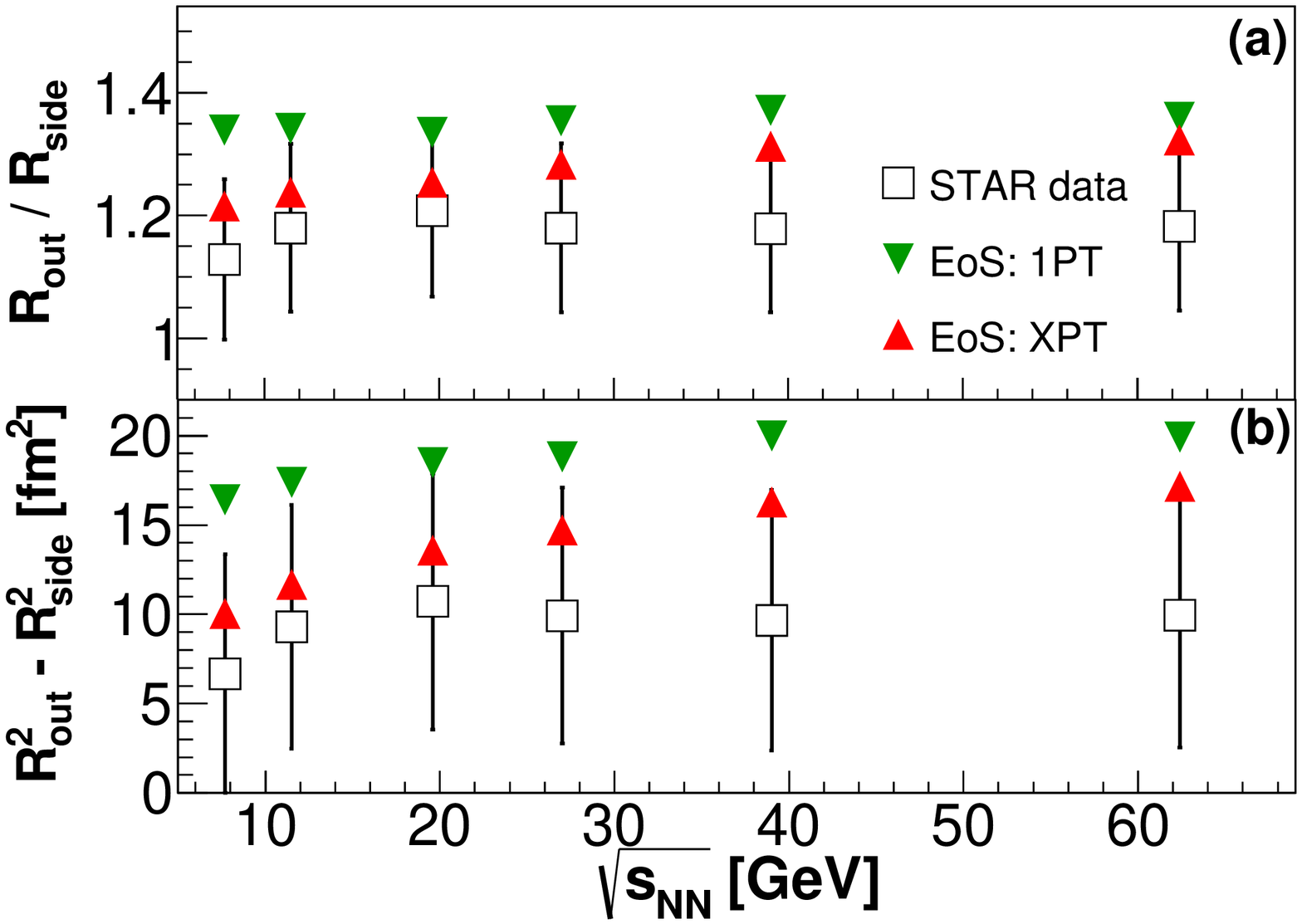}
    \caption{Ratio of the ``out'' and ``side'' radii (a) and difference of the radii squared (b) as a function of $\sqrt{s_{NN}}$ derived from the STAR data ($0.15 < k_{T} < 0.25$ GeV/c, 0-5\% centrality) and compared with the model calculations using the two EoS's.}  
    \label{fig:Rdiff}
  \end{center}
\end{figure}

The $R_{out} / R_{side}$ and $R_{out}^{2} - R_{side}^{2}$ as a function of $\sqrt{s_{NN}}$ were studied at fixed $m_{T}$ by the STAR collaboration~\cite{STAR_HBT}. A wide maximum near $\sqrt{s_{NN}}$ $\sim$~20~GeV/c in both excitation functions was observed. This observation is however accompanied by rather large systematic error bars. We have calculated the very same quantities in the model and compared them with experimental data. The result of comparison is shown in Fig.~\ref{fig:Rdiff}.

One can see that due to large experimental error bars the model calculations involving the XPT EoS are in a strong agreement with the data within the error bars at all energies, whereas the 1PT EoS overestimates the data. However, in the model taking into account the XPT EoS we observe a monotonic increase in excitation functions of both quantities, meanwhile the 1PT EoS results in a non-decreasing behavior of the quantities. The XPT EoS ``works'' better for lowest collision energies that might be seen earlier from a better description of individual radii in that energy region shown in Fig.~\ref{fig:R3Dall}. A study of the $R_{out} / R_{side}$ ratio looks traditional in the modern femtoscopy since the $R_{out}$ and $R_{side}$ radii are both reduced by flow, thus their ratio is a more robust against the flow effects.

As mentioned above, the parameters of the model were adjusted to approach the basic hadronic observables: rapidity, transverse momentum distributions and elliptic flow coefficients within the BES region, but not femtoscopic ones. No model tuning has been made for the femtoscopy, therefore the obtained radii may be considered as a free ``prediction'' even though the experimental data already exists.

\subsection{Source emission functions}
In a Monte Carlo model one has an access to the space-time characteristics of produced particles, which allows one to avoid the complicated procedure of solving the integral equation (see Eq.~\ref{Koonin_Pratt}) as it is done in experiment~\cite{SF}. The source emission function can be calculated directly as:

\begin{equation}
S(r^{*})=\frac{\sum_{i\ne j} \delta_\Delta (\bf{r}^{*}-\bf{r}_{i}^{*}+\bf{r}_{j}^{*})}{N \Delta^3}
\label{Snorm}
\end{equation}

Here $r^{∗}_i$ and $r^{∗}_j$ are the particles space positions, $r^{∗}$ is the particles separation in PRF;  $\delta_{\Delta}$ = 1 if $|x| < p/2$ and 0 otherwise, $p$ is a size of the histogram bin. The denominator in Eq.~(\ref{Snorm}) takes care for the normalization by a product of the number of pairs $N = \sum_{i \ne j} 1$ and a bin size $\Delta^3$.

\begin{figure*}[!th]
  \begin{center}
    \includegraphics[width=1.0\textwidth]{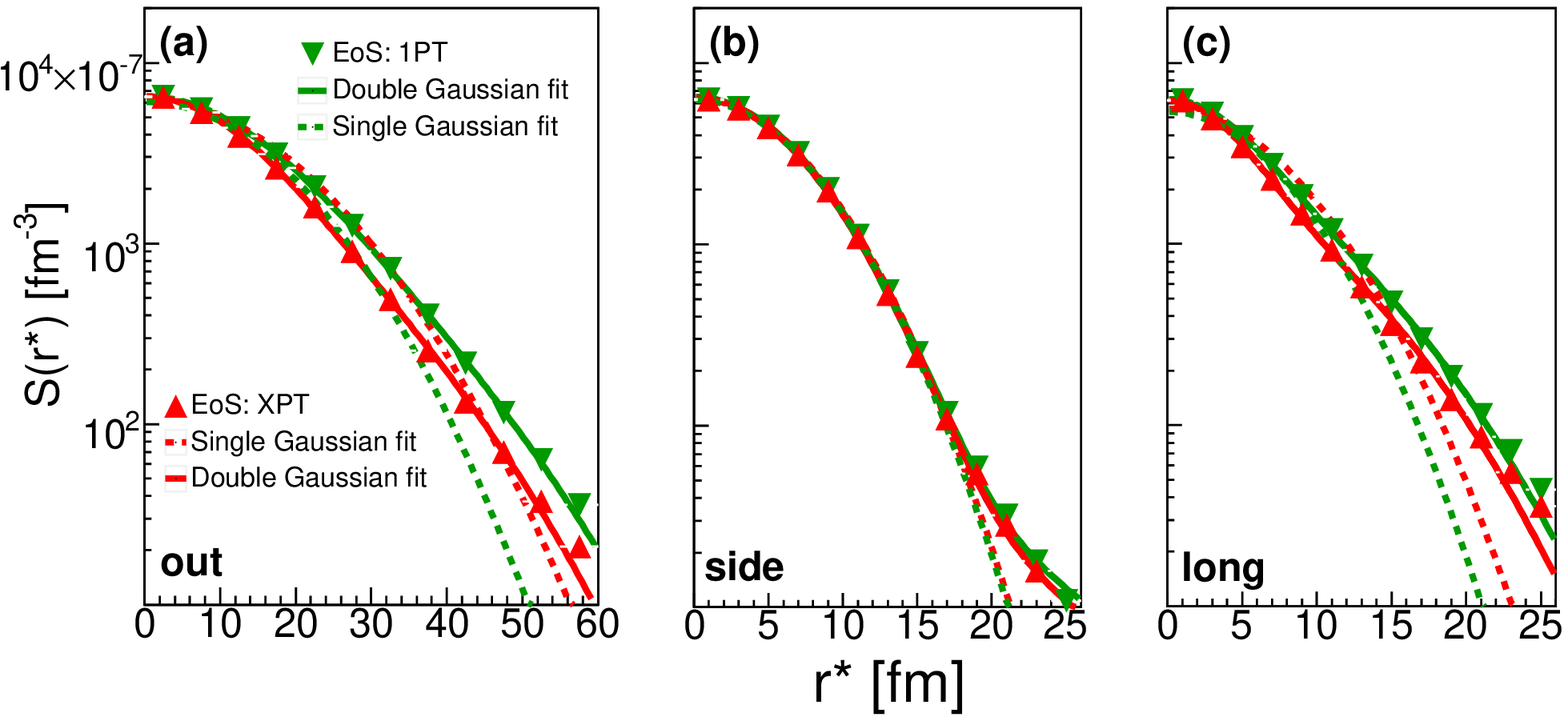}
    \caption{
    One-dimensional projections of source emission functions of pions from the model (full simulation with cascade) obtained at $\sqrt{s_{NN}}$ = 7.7~GeV for pion pairs satisfying the cut on transverse momentum: 0.2$ < k_{T} < $0.4 GeV/c.}   
    \label{fig:SF}
  \end{center}
\end{figure*}

Fig.~\ref{fig:SF} demonstrates an example of one-dimensional projections of source emission function $S(r^{*})$ derived from the model directly. 

One can see that calculations involving the 1PT EoS lead to longer visible tails in the projections as compared with the XPT EoS, especially for ``out'' direction. It is related to a weaker transverse flow developed in the fluid phase and a longer lifetime of the fluid phase taking place in the 1PT EoS. A similar observation has been reported for the ``out'' femtoscopic radii in the previous section.

A set of functions consisting of a single Gaussian, double Gaussian, Gaussian + Exponential, Gaussian + Lorentzian, and Hump function~\cite{Alt:HumpFunction} was tested for a description of one-dimensional projections of source emission functions. The best description was obtained with the  Hump-function and the double Gaussian function. The last one gives only slightly worse $\chi^2$ than the Hump-function, but allows one for a clear interpretation of parameters and a more stable fit. The single Gaussian and double Gaussian fit functions are shown in Fig.~\ref{fig:SF}. The parameters extracted from these fits are presented in Table~\ref{tab:SF_fits}.

\begin{table*}[!th]
\caption{Results of single (Eq.~(\ref{SF_Gaussian})) and double Gaussian fits of model source emission functions shown in Fig.~\ref{fig:SF}. $\chi^{2} / ndf$ values in parenthesis correspond to the XPT EoS. All calculations are performed in PRF.}

\begin{center}
\begin{tabular}{|c|c|c|c|c|c|}\hline
Single Gaussian & $\chi^2/ndf$ & Radius & 1PT & XPT \\ 
\hline
           &   975 (1247)  &  $R_{out}$ [fm] & 11.10 $\pm$ 0.01  &  10.03 $\pm$ 0.01 \\
 7.7 GeV   &   300 (251)  &  $R_{side}$ [fm] & 4.19 $\pm$ 0.01  &  4.16  $\pm$ 0.01 \\
           &   3259 (3878)  &  $R_{long} [fm] $ & 4.59 $\pm$ 0.01  &  4.20 $\pm$ 0.01 \\
\hline
\hline
Double Gaussian & $\chi^{2}/ndf$ & Radius & 1PT & XPT \\ 
\hline
           &    51.6 (73.0) &  $R_{out}^{small}$ [fm] & 8.91 $\pm$ 0.01  ($\lambda_{out}^{small}$=0.66) &  7.83 $\pm$ 0.01 ($\lambda_{out}^{small}$=0.68)\\
 7.7 GeV   &    58.7 (66.6) &  $R_{side}^{small}$ [fm] & 4.10 $\pm$ 0.01  ($\lambda_{side}^{small}$=0.99)  &  4.08  $\pm$ 0.01 ($\lambda_{side}^{small}$=0.99) \\
           &    130.3 (195.1) &  $R_{long}^{small}$ [fm] & 3.10 $\pm$ 0.01 ($\lambda_{long}^{small}$=0.64)  &  2.93 $\pm$ 0.01 ($\lambda_{long}^{small}$=0.74)\\
\hline
           &               &  $R_{out}^{large}$ [fm] & 13.85 $\pm$ 0.01 ($\lambda_{out}^{large}$=0.34)    &  12.88 $\pm$ 0.01 ($\lambda_{out}^{large}$=0.32)  \\
           &               &  $R_{side}^{large}$ [fm] & 9.76 $\pm$ 0.01 ($\lambda_{side}^{large}$=0.01)   &  9.45  $\pm$ 0.01 ($\lambda_{side}^{large}$=0.01) \\
           &               &  $R_{long}^{large}$ [fm] & 6.06 $\pm$ 0.01  ($\lambda_{long}^{large}$=0.36)   &  6.24 $\pm$ 0.01 ($\lambda_{long}^{large}$=0.26) \\
\hline
\end{tabular}
\end{center}
\label{tab:SF_fits}
\end{table*}

The fit of projections of source emission function with a single Gaussian gives a large value of $\chi^2/ndf$. The fit with a double Gaussian allows one to get much better values of $\chi^2/ndf$ and obtain a better description of the tails of projections of source emission functions until $\sim$~60~fm in ``out'' and $\sim$~25~fm in ``side'' and ``long'' directions, respectively. The radii extracted from the double Gaussian fit have a small component $R_{i}^{small}$ of 4-12~fm and a large component $R_{i}^{large}$ of 8-20~fm (as usual, $i$ denotes ``out'', ``side'' and ``long'' directions). It reflects the fact that pion source consists of direct particles (described by the first component) and re-scatterings (the second one). The difference between radii extracted from the source emission functions obtained with the two EoS's is seen for both components - $R_{i}^{small}$ and $R_{i}^{large}$, but it is rather small, less than 0.5~fm. The radii are larger for the 1PT scenario being consistent with the three-dimensional femtoscopic radii reported above. 

It is interesting to note that in case of the single Gaussian fit the values of the radii are approximately equal to the ones derived from the double Gaussian fit and averaged quadratically over relative contributions of small and large radii. It means that 
the one-dimensional Gaussian radii roughly reflect the main features of double Gaussian fits.

\begin{figure*}[!th]
  \begin{center}
    \includegraphics[width=1.0\textwidth]{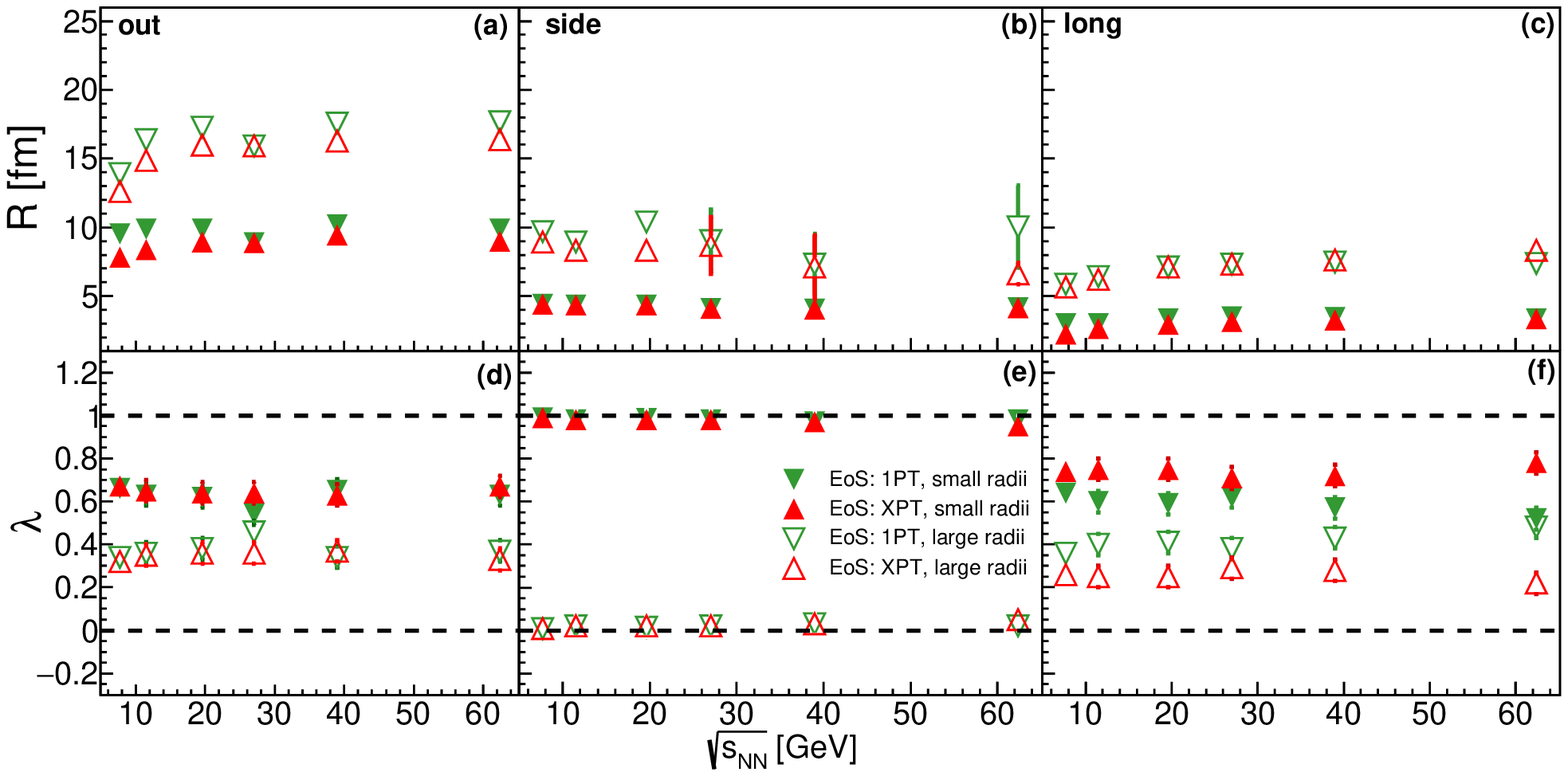}
    \caption{$\sqrt{s_{NN}}$-dependence of radii extracted from the double Gaussian fit ((a) - (c)) and relative contribution of them ($\lambda$) to ``out'', ``side'' and ``long'' directions ((d) - (f)). Pion pairs satisfying 0.2$ < k_{T} < $0.8~GeV/c are taken into account.}
    \label{fig:SFsNN}
  \end{center}
\end{figure*}

Fig.~\ref{fig:SFsNN} shows a $\sqrt{s_{NN}}$-dependence of the small and large radii and their relative contributions extracted from the double Gaussian fit depending on EoS.

The radii increase with increasing $\sqrt{s_{NN}}$ for both types of calculations. The visible difference between small and large  radii in ``out'' direction (see Fig.\ref{fig:SFsNN} (a)) decreases with increase of $\sqrt{s_{NN}}$. The relative contributions of small and large radii to ``out'' direction are equal to $\sim 0.65$ and $\sim 0.35$ (see Fig.\ref{fig:SFsNN} (d)) and practically do not depend on either $\sqrt{s_{NN}}$ or type of EoS. 
The radii in ``side'' direction seem to be independent of $\sqrt{s_{NN}}$ (see Fig.\ref{fig:SFsNN} (b), (e)). The radii in ``long'' projection almost coincide for both types of EoS (see Fig.\ref{fig:SFsNN} (c)), but the relative contributions of them  as a function of $\sqrt{s_{NN}}$ demonstrate a difference depending on EoS. The relative contribution of the large radii has a tendency to increase with $\sqrt{s_{NN}}$ and is larger in case of the 1PT scenario (see Fig.\ref{fig:SFsNN} (f)).

Of course, the best comparison with experiment is a direct comparison of source emission functions from the model with the extracted ones experimentally. Nevertheless, this study shows us that the use of a double Gaussian fit also reflects a lot of interesting features of source emission functions, while a single Gaussian fit used in many experiments can be a rather risky procedure due to poor description of the source by this function. 
However, as it was demonstrated above, the radii extracted from the single Gaussian fit are equal to the properly averaged double Gaussian radii, giving, in principle, a realistic information on the source. This result is quite encouraging since it is much easier to study the three-dimensional radii than the source emission functions.

\section{Summary}
\label{sec:sec4}
We have presented the first study of pion femtoscopy in viscous hydro + cascade model \texttt{vHLLE+UrQMD} in the energy range of the BES program at RHIC.
It is shown that the chiral model EoS \cite{SchrammSteinheimer} (XPT EoS), which has a crossover-type transition between QGP and hadron gas phases, in the fluid phase results in a quite reasonable reproduction of three-dimensional pion femtoscopic radii measured by the STAR collaboration. 

The ``out'' Gaussian femtoscopic radii obtained with the bag model EoS (1PT EoS) are systematically larger as compared with the XPT EoS; the ``side'' radii coincide for both types of EoS; the ``long'' radii are also somewhat larger for the 1PT EoS.
 
The 1PT EoS results in a systematically worse reproduction of the data, however the differences between two EoS's are not so large.
The $R_{out} / R_{side}$ ratio and $R_{out}^2 - R_{side}^2$  are in agreement with the STAR results within the error bars at all collision energies using the XPT EoS, but their energy dependences observed in the model are quite monotonic as opposed to the broad maximum around $\sqrt{s_{NN}} = 20$~GeV reported by STAR. At the same time, the 1PT EoS overestimates experimental data points for both $R_{out} / R_{side}$ and $R_{out}^2 - R_{side}^2$.  In particular, the latter EoS  does not reproduce the femtoscopic radii even at the lowest energy considered, $\sqrt{s_{NN}} = 7.7$~GeV.

Parameters of the model were adjusted in \cite{vHLLE_UrQMD_Yura1} based on rapidity, transverse momentum spectra and elliptic flow data in the BES region for the XPT EoS scenario. No readjustment for the 1PT EoS has been made, which poses an open question whether the differences in femtoscopic radii between the two EoS will be even smaller if the readjustment is made for each EoS scenario individually. Also, no additional parameter tuning has been made for the femtoscopic observables, therefore the results may be considered as ``free model predictions'' even though the experimental data already exists.

We find that a better overall description of the femtoscopic radii would require about 1~fm/c shorter duration of pion emission with the present setup of the model. If it is realized, then at lower energies the 1PT scenario will be closer to the data. This in turn may indicate a change of the nature of phase transition at energies less than $\sqrt{s_{\rm NN}}=20$~GeV and this should be an incentive for future experiments at NICA and FAIR facilities. It is an open question whether a new set of parameters more suitable for the femtoscopic radii description can be found.

In addition to traditional femtoscopic radii, we have calculated source emission functions of pion pairs. We show that it is possible to distinguish calculations with the two different EoS. The projections of source emission functions onto ``out'' direction are wider for the use of the 1PT EoS. For ``side'' direction these projections coincide for both scenarios; for ``long'' direction the projections obtained with the 1PT EoS are also wider in comparison with calculations using the XPT EoS. This observation is related to a weaker transverse flow developed in the fluid phase and a longer lifetime of the phase in case of the 1PT EoS used. 

In order to describe the source emission functions quantitatively a set of different fitting functions has been tested.
It is shown that the use of a double Gaussian fit to the source emission function gives a reasonable description and allows one for a simple interpretation of the obtained small and large radii.

So far we have performed femtoscopic analysis with vHLLE+UrQMD model only. As a next step we plan to extend the analysis using 3-fluid hydrodynamics-based event generator THESEUS~\cite{Batyuk:THESEUS}. In THESEUS the hydrodynamical description of heavy ion reaction starts earlier, which results in different sensitivity to hydrodynamic EoS especially in the NICA energy range.

\addcontentsline{toc}{section}{\bibname}
\bibliography{biblio.bib}

\begin{thebibliography}{56}%
\makeatletter
\providecommand \@ifxundefined [1]{%
 \@ifx{#1\undefined}
}%
\providecommand \@ifnum [1]{%
 \ifnum #1\expandafter \@firstoftwo
 \else \expandafter \@secondoftwo
 \fi
}%
\providecommand \@ifx [1]{%
 \ifx #1\expandafter \@firstoftwo
 \else \expandafter \@secondoftwo
 \fi
}%
\providecommand \natexlab [1]{#1}%
\providecommand \enquote  [1]{``#1''}%
\providecommand \bibnamefont  [1]{#1}%
\providecommand \bibfnamefont [1]{#1}%
\providecommand \citenamefont [1]{#1}%
\providecommand \href@noop [0]{\@secondoftwo}%
\providecommand \href [0]{\begingroup \@sanitize@url \@href}%
\providecommand \@href[1]{\@@startlink{#1}\@@href}%
\providecommand \@@href[1]{\endgroup#1\@@endlink}%
\providecommand \@sanitize@url [0]{\catcode `\\12\catcode `\$12\catcode
  `\&12\catcode `\#12\catcode `\^12\catcode `\_12\catcode `\%12\relax}%
\providecommand \@@startlink[1]{}%
\providecommand \@@endlink[0]{}%
\providecommand \url  [0]{\begingroup\@sanitize@url \@url }%
\providecommand \@url [1]{\endgroup\@href {#1}{\urlprefix }}%
\providecommand \urlprefix  [0]{URL }%
\providecommand \Eprint [0]{\href }%
\providecommand \doibase [0]{http://dx.doi.org/}%
\providecommand \selectlanguage [0]{\@gobble}%
\providecommand \bibinfo  [0]{\@secondoftwo}%
\providecommand \bibfield  [0]{\@secondoftwo}%
\providecommand \translation [1]{[#1]}%
\providecommand \BibitemOpen [0]{}%
\providecommand \bibitemStop [0]{}%
\providecommand \bibitemNoStop [0]{.\EOS\space}%
\providecommand \EOS [0]{\spacefactor3000\relax}%
\providecommand \BibitemShut  [1]{\csname bibitem#1\endcsname}%
\let\auto@bib@innerbib\@empty
\bibitem [{\citenamefont {Cabibbo}\ and\ \citenamefont {Parisi}(1975)}]{QGP_a}%
  \BibitemOpen
  \bibfield  {author} {\bibinfo {author} {\bibfnamefont {N.}~\bibnamefont
  {Cabibbo}}\ and\ \bibinfo {author} {\bibfnamefont {G.}~\bibnamefont
  {Parisi}},\ }\href@noop {} {\bibfield  {journal} {\bibinfo  {journal} {Phys.
  Lett.B}\ }\textbf {\bibinfo {volume} {59}},\ \bibinfo {pages} {67} (\bibinfo
  {year} {1975})}\BibitemShut {NoStop}%
\bibitem [{\citenamefont {Shuryak}(1980)}]{QGP_b}%
  \BibitemOpen
  \bibfield  {author} {\bibinfo {author} {\bibfnamefont {E.~V.}\ \bibnamefont
  {Shuryak}},\ }\href@noop {} {\bibfield  {journal} {\bibinfo  {journal}
  {Physics Reports}\ }\textbf {\bibinfo {volume} {61}},\ \bibinfo {pages} {71}
  (\bibinfo {year} {1980})}\BibitemShut {NoStop}%
\bibitem [{\citenamefont {Arsene}\ \emph {et~al.}(2005)\citenamefont {Arsene}
  \emph {et~al.}}]{RHIC_hydro_a}%
  \BibitemOpen
  \bibfield  {author} {\bibinfo {author} {\bibfnamefont {I.}~\bibnamefont
  {Arsene}} \emph {et~al.},\ }\href@noop {} {\bibfield  {journal} {\bibinfo
  {journal} {Nucl. Phys. A}\ }\textbf {\bibinfo {volume} {757}},\ \bibinfo
  {pages} {1 } (\bibinfo {year} {2005})}\BibitemShut {NoStop}%
\bibitem [{\citenamefont {Back}\ \emph {et~al.}(2005)\citenamefont {Back} \emph
  {et~al.}}]{RHIC_hydro_b}%
  \BibitemOpen
  \bibfield  {author} {\bibinfo {author} {\bibfnamefont {B.}~\bibnamefont
  {Back}} \emph {et~al.},\ }\href@noop {} {\bibfield  {journal} {\bibinfo
  {journal} {Nucl. Phys. A}\ }\textbf {\bibinfo {volume} {757}},\ \bibinfo
  {pages} {28 } (\bibinfo {year} {2005})}\BibitemShut {NoStop}%
\bibitem [{\citenamefont {Adams}\ \emph {et~al.}(2005)\citenamefont {Adams}
  \emph {et~al.}}]{RHIC_hydro_c}%
  \BibitemOpen
  \bibfield  {author} {\bibinfo {author} {\bibfnamefont {J.}~\bibnamefont
  {Adams}} \emph {et~al.},\ }\href@noop {} {\bibfield  {journal} {\bibinfo
  {journal} {Nucl. Phys. A}\ }\textbf {\bibinfo {volume} {757}},\ \bibinfo
  {pages} {102 } (\bibinfo {year} {2005})}\BibitemShut {NoStop}%
\bibitem [{\citenamefont {Adcox}\ \emph {et~al.}(2005)\citenamefont {Adcox}
  \emph {et~al.}}]{RHIC_hydro_d}%
  \BibitemOpen
  \bibfield  {author} {\bibinfo {author} {\bibfnamefont {K.}~\bibnamefont
  {Adcox}} \emph {et~al.},\ }\href@noop {} {\bibfield  {journal} {\bibinfo
  {journal} {Nucl. Phys. A}\ }\textbf {\bibinfo {volume} {757}},\ \bibinfo
  {pages} {184 } (\bibinfo {year} {2005})}\BibitemShut {NoStop}%
\bibitem [{\citenamefont {Aamodt}\ \emph {et~al.}(2010)\citenamefont {Aamodt}
  \emph {et~al.}}]{ALICE_hydro_a}%
  \BibitemOpen
  \bibfield  {author} {\bibinfo {author} {\bibfnamefont {K.}~\bibnamefont
  {Aamodt}} \emph {et~al.},\ }\href@noop {} {\bibfield  {journal} {\bibinfo
  {journal} {Phys. Rev. Lett.}\ }\textbf {\bibinfo {volume} {105}},\ \bibinfo
  {pages} {252302} (\bibinfo {year} {2010})}\BibitemShut {NoStop}%
\bibitem [{\citenamefont {Abelev}\ \emph {et~al.}(2015)\citenamefont {Abelev}
  \emph {et~al.}}]{ALICE_hydro_b}%
  \BibitemOpen
  \bibfield  {author} {\bibinfo {author} {\bibnamefont {Abelev}} \emph
  {et~al.},\ }\href@noop {} {\bibfield  {journal} {\bibinfo  {journal} {JHEP}\
  }\textbf {\bibinfo {volume} {06}},\ \bibinfo {pages} {190} (\bibinfo {year}
  {2015})}\BibitemShut {NoStop}%
\bibitem [{\citenamefont {Adam}\ \emph {et~al.}(2015)\citenamefont {Adam} \emph
  {et~al.}}]{ALICE_hydro_c}%
  \BibitemOpen
  \bibfield  {author} {\bibinfo {author} {\bibfnamefont {J.}~\bibnamefont
  {Adam}} \emph {et~al.},\ }\href@noop {} {\bibfield  {journal} {\bibinfo
  {journal} {Physics Letters B}\ }\textbf {\bibinfo {volume} {746}},\ \bibinfo
  {pages} {1 } (\bibinfo {year} {2015})}\BibitemShut {NoStop}%
\bibitem [{\citenamefont {Aamodt}\ \emph {et~al.}(2011)\citenamefont {Aamodt}
  \emph {et~al.}}]{ALICE_hydro_d}%
  \BibitemOpen
  \bibfield  {author} {\bibinfo {author} {\bibfnamefont {K.}~\bibnamefont
  {Aamodt}} \emph {et~al.},\ }\href@noop {} {\bibfield  {journal} {\bibinfo
  {journal} {Phys. Lett. B}\ }\textbf {\bibinfo {volume} {696}},\ \bibinfo
  {pages} {328} (\bibinfo {year} {2011})}\BibitemShut {NoStop}%
\bibitem [{\citenamefont {Sinyukov}\ \emph {et~al.}(2009)\citenamefont
  {Sinyukov}, \citenamefont {Akkelin}, \citenamefont {Karpenko},\ and\
  \citenamefont {Hama}}]{Sin09}%
  \BibitemOpen
  \bibfield  {author} {\bibinfo {author} {\bibfnamefont {Y.~M.}\ \bibnamefont
  {Sinyukov}}, \bibinfo {author} {\bibfnamefont {S.~V.}\ \bibnamefont
  {Akkelin}}, \bibinfo {author} {\bibfnamefont {I.~A.}\ \bibnamefont
  {Karpenko}}, \ and\ \bibinfo {author} {\bibfnamefont {Y.}~\bibnamefont
  {Hama}},\ }\href@noop {} {\bibfield  {journal} {\bibinfo  {journal} {Acta
  Phys. Polon. B}\ }\textbf {\bibinfo {volume} {40}},\ \bibinfo {pages} {1025 }
  (\bibinfo {year} {2009})}\BibitemShut {NoStop}%
\bibitem [{\citenamefont {Broniowski}\ \emph {et~al.}(2008)\citenamefont
  {Broniowski}, \citenamefont {Chojnacki}, \citenamefont {Florkowski},\ and\
  \citenamefont {Kisiel}}]{AK08}%
  \BibitemOpen
  \bibfield  {author} {\bibinfo {author} {\bibfnamefont {W.}~\bibnamefont
  {Broniowski}}, \bibinfo {author} {\bibfnamefont {M.}~\bibnamefont
  {Chojnacki}}, \bibinfo {author} {\bibfnamefont {W.}~\bibnamefont
  {Florkowski}}, \ and\ \bibinfo {author} {\bibfnamefont {A.}~\bibnamefont
  {Kisiel}},\ }\href@noop {} {\bibfield  {journal} {\bibinfo  {journal} {Phys.
  Rev. Lett.}\ }\textbf {\bibinfo {volume} {101}},\ \bibinfo {pages} {022301}
  (\bibinfo {year} {2008})}\BibitemShut {NoStop}%
\bibitem [{\citenamefont {Pratt}(2009)}]{Pratt09}%
  \BibitemOpen
  \bibfield  {author} {\bibinfo {author} {\bibfnamefont {S.}~\bibnamefont
  {Pratt}},\ }\href@noop {} {\bibfield  {journal} {\bibinfo  {journal} {Nucl.
  Phys. A}\ }\textbf {\bibinfo {volume} {830}},\ \bibinfo {pages} {51C}
  (\bibinfo {year} {2009})}\BibitemShut {NoStop}%
\bibitem [{\citenamefont {Werner}\ \emph {et~al.}(2010)\citenamefont {Werner},
  \citenamefont {Karpenko}, \citenamefont {Pierog}, \citenamefont {Bleicher},\
  and\ \citenamefont {Mikhailov}}]{Werner:2010aa}%
  \BibitemOpen
  \bibfield  {author} {\bibinfo {author} {\bibfnamefont {K.}~\bibnamefont
  {Werner}}, \bibinfo {author} {\bibfnamefont {I.}~\bibnamefont {Karpenko}},
  \bibinfo {author} {\bibfnamefont {T.}~\bibnamefont {Pierog}}, \bibinfo
  {author} {\bibfnamefont {M.}~\bibnamefont {Bleicher}}, \ and\ \bibinfo
  {author} {\bibfnamefont {K.}~\bibnamefont {Mikhailov}},\ }\href@noop {}
  {\bibfield  {journal} {\bibinfo  {journal} {Phys. Rev. C}\ }\textbf {\bibinfo
  {volume} {82}},\ \bibinfo {pages} {044904} (\bibinfo {year}
  {2010})}\BibitemShut {NoStop}%
\bibitem [{\citenamefont {Bo\ifmmode~\dot{z}\else
  \.{z}\fi{}ek}(2011)}]{Bosek_2011}%
  \BibitemOpen
  \bibfield  {author} {\bibinfo {author} {\bibfnamefont {P.}~\bibnamefont
  {Bo\ifmmode~\dot{z}\else \.{z}\fi{}ek}},\ }\href@noop {} {\bibfield
  {journal} {\bibinfo  {journal} {Phys. Rev. C}\ }\textbf {\bibinfo {volume}
  {83}},\ \bibinfo {pages} {044910} (\bibinfo {year} {2011})}\BibitemShut
  {NoStop}%
\bibitem [{\citenamefont {Karpenko}\ \emph {et~al.}(2013)\citenamefont
  {Karpenko}, \citenamefont {Sinyukov},\ and\ \citenamefont
  {Werner}}]{Karpenko:2012yf}%
  \BibitemOpen
  \bibfield  {author} {\bibinfo {author} {\bibfnamefont {I.~A.}\ \bibnamefont
  {Karpenko}}, \bibinfo {author} {\bibfnamefont {{\relax Yu}.~M.}\ \bibnamefont
  {Sinyukov}}, \ and\ \bibinfo {author} {\bibfnamefont {K.}~\bibnamefont
  {Werner}},\ }\href@noop {} {\bibfield  {journal} {\bibinfo  {journal} {Phys.
  Rev. C}\ }\textbf {\bibinfo {volume} {87}},\ \bibinfo {pages} {024914}
  (\bibinfo {year} {2013})}\BibitemShut {NoStop}%
\bibitem [{\citenamefont {Brown}\ \emph {et~al.}(1990)\citenamefont {Brown},
  \citenamefont {Butler}, \citenamefont {Chen}, \citenamefont {Christ},
  \citenamefont {Dong}, \citenamefont {Schaffer}, \citenamefont {Unger},\ and\
  \citenamefont {Vaccarino}}]{Lat6_a}%
  \BibitemOpen
  \bibfield  {author} {\bibinfo {author} {\bibfnamefont {F.~R.}\ \bibnamefont
  {Brown}}, \bibinfo {author} {\bibfnamefont {F.~P.}\ \bibnamefont {Butler}},
  \bibinfo {author} {\bibfnamefont {H.}~\bibnamefont {Chen}}, \bibinfo {author}
  {\bibfnamefont {N.~H.}\ \bibnamefont {Christ}}, \bibinfo {author}
  {\bibfnamefont {Z.}~\bibnamefont {Dong}}, \bibinfo {author} {\bibfnamefont
  {W.}~\bibnamefont {Schaffer}}, \bibinfo {author} {\bibfnamefont {L.~I.}\
  \bibnamefont {Unger}}, \ and\ \bibinfo {author} {\bibfnamefont
  {A.}~\bibnamefont {Vaccarino}},\ }\href@noop {} {\bibfield  {journal}
  {\bibinfo  {journal} {Phys. Rev. Lett.}\ }\textbf {\bibinfo {volume} {65}},\
  \bibinfo {pages} {2491} (\bibinfo {year} {1990})}\BibitemShut {NoStop}%
\bibitem [{\citenamefont {Aoki}\ \emph {et~al.}(2006)\citenamefont {Aoki} \emph
  {et~al.}}]{Lat6_b}%
  \BibitemOpen
  \bibfield  {author} {\bibinfo {author} {\bibfnamefont {Y.}~\bibnamefont
  {Aoki}} \emph {et~al.},\ }\href@noop {} {\bibfield  {journal} {\bibinfo
  {journal} {Nature}\ }\textbf {\bibinfo {volume} {443}},\ \bibinfo {pages}
  {675} (\bibinfo {year} {2006})}\BibitemShut {NoStop}%
\bibitem [{\citenamefont {Borsanyi}\ \emph {et~al.}(2010)\citenamefont
  {Borsanyi}, \citenamefont {Fodor}, \citenamefont {Hoelbling}, \citenamefont
  {Katz}, \citenamefont {Krieg}, \citenamefont {Ratti},\ and\ \citenamefont
  {Szabo}}]{Lat6_c}%
  \BibitemOpen
  \bibfield  {author} {\bibinfo {author} {\bibfnamefont {S.}~\bibnamefont
  {Borsanyi}}, \bibinfo {author} {\bibfnamefont {Z.}~\bibnamefont {Fodor}},
  \bibinfo {author} {\bibfnamefont {C.}~\bibnamefont {Hoelbling}}, \bibinfo
  {author} {\bibfnamefont {S.~D.}\ \bibnamefont {Katz}}, \bibinfo {author}
  {\bibfnamefont {S.}~\bibnamefont {Krieg}}, \bibinfo {author} {\bibfnamefont
  {C.}~\bibnamefont {Ratti}}, \ and\ \bibinfo {author} {\bibfnamefont {K.~K.}\
  \bibnamefont {Szabo}},\ }\href@noop {} {\bibfield  {journal} {\bibinfo
  {journal} {JHEP}\ }\textbf {\bibinfo {volume} {09}},\ \bibinfo {pages} {073}
  (\bibinfo {year} {2010})}\BibitemShut {NoStop}%
\bibitem [{\citenamefont {Cheng}(2009)}]{Lat6_d}%
  \BibitemOpen
  \bibfield  {author} {\bibinfo {author} {\bibfnamefont {M.}~\bibnamefont
  {Cheng}},\ }\href@noop {} {\bibfield  {journal} {\bibinfo  {journal} {PoS}\
  }\textbf {\bibinfo {volume} {LAT2009}},\ \bibinfo {pages} {175} (\bibinfo
  {year} {2009})}\BibitemShut {NoStop}%
\bibitem [{\citenamefont {Karsch}\ \emph {et~al.}(2016)\citenamefont {Karsch}
  \emph {et~al.}}]{Karsch:2015nqx}%
  \BibitemOpen
  \bibfield  {author} {\bibinfo {author} {\bibfnamefont {F.}~\bibnamefont
  {Karsch}} \emph {et~al.},\ }\href@noop {} {\bibfield  {journal} {\bibinfo
  {journal} {Nucl. Phys. A}\ }\textbf {\bibinfo {volume} {956}},\ \bibinfo
  {pages} {352} (\bibinfo {year} {2016})}\BibitemShut {NoStop}%
\bibitem [{\citenamefont {Pratt}\ \emph {et~al.}(2015)\citenamefont {Pratt},
  \citenamefont {Sangaline}, \citenamefont {Sorensen},\ and\ \citenamefont
  {Wang}}]{Pratt:2015zsa}%
  \BibitemOpen
  \bibfield  {author} {\bibinfo {author} {\bibfnamefont {S.}~\bibnamefont
  {Pratt}}, \bibinfo {author} {\bibfnamefont {E.}~\bibnamefont {Sangaline}},
  \bibinfo {author} {\bibfnamefont {P.}~\bibnamefont {Sorensen}}, \ and\
  \bibinfo {author} {\bibfnamefont {H.}~\bibnamefont {Wang}},\ }\href@noop {}
  {\bibfield  {journal} {\bibinfo  {journal} {Phys. Rev. Lett.}\ }\textbf
  {\bibinfo {volume} {114}},\ \bibinfo {pages} {202301} (\bibinfo {year}
  {2015})}\BibitemShut {NoStop}%
\bibitem [{\citenamefont {Masayuki}\ and\ \citenamefont
  {Koichi}(1989)}]{ML10_16_a}%
  \BibitemOpen
  \bibfield  {author} {\bibinfo {author} {\bibfnamefont {A.}~\bibnamefont
  {Masayuki}}\ and\ \bibinfo {author} {\bibfnamefont {Y.}~\bibnamefont
  {Koichi}},\ }\href@noop {} {\bibfield  {journal} {\bibinfo  {journal} {Nucl.
  Phys. A}\ }\textbf {\bibinfo {volume} {504}},\ \bibinfo {pages} {668 }
  (\bibinfo {year} {1989})}\BibitemShut {NoStop}%
\bibitem [{\citenamefont {Barducci}\ \emph {et~al.}(1994)\citenamefont
  {Barducci}, \citenamefont {Casalbuoni}, \citenamefont {Pettini},\ and\
  \citenamefont {Gatto}}]{ML10_16_b}%
  \BibitemOpen
  \bibfield  {author} {\bibinfo {author} {\bibfnamefont {A.}~\bibnamefont
  {Barducci}}, \bibinfo {author} {\bibfnamefont {R.}~\bibnamefont
  {Casalbuoni}}, \bibinfo {author} {\bibfnamefont {G.}~\bibnamefont {Pettini}},
  \ and\ \bibinfo {author} {\bibfnamefont {R.}~\bibnamefont {Gatto}},\
  }\href@noop {} {\bibfield  {journal} {\bibinfo  {journal} {Phys. Rev. D}\
  }\textbf {\bibinfo {volume} {49}},\ \bibinfo {pages} {426} (\bibinfo {year}
  {1994})}\BibitemShut {NoStop}%
\bibitem [{\citenamefont {Berges}\ and\ \citenamefont
  {Rajagopal}(1999)}]{ML10_16_c}%
  \BibitemOpen
  \bibfield  {author} {\bibinfo {author} {\bibfnamefont {J.}~\bibnamefont
  {Berges}}\ and\ \bibinfo {author} {\bibfnamefont {K.}~\bibnamefont
  {Rajagopal}},\ }\href@noop {} {\bibfield  {journal} {\bibinfo  {journal}
  {Nuclear Physics B}\ }\textbf {\bibinfo {volume} {538}},\ \bibinfo {pages}
  {215 } (\bibinfo {year} {1999})}\BibitemShut {NoStop}%
\bibitem [{\citenamefont {Halasz}\ \emph {et~al.}(1998)\citenamefont {Halasz},
  \citenamefont {Jackson}, \citenamefont {Shrock}, \citenamefont {Stephanov},\
  and\ \citenamefont {Verbaarschot}}]{ML10_16_d}%
  \BibitemOpen
  \bibfield  {author} {\bibinfo {author} {\bibfnamefont {M.~A.}\ \bibnamefont
  {Halasz}}, \bibinfo {author} {\bibfnamefont {A.~D.}\ \bibnamefont {Jackson}},
  \bibinfo {author} {\bibfnamefont {R.~E.}\ \bibnamefont {Shrock}}, \bibinfo
  {author} {\bibfnamefont {M.~A.}\ \bibnamefont {Stephanov}}, \ and\ \bibinfo
  {author} {\bibfnamefont {J.~J.~M.}\ \bibnamefont {Verbaarschot}},\
  }\href@noop {} {\bibfield  {journal} {\bibinfo  {journal} {Phys. Rev. D}\
  }\textbf {\bibinfo {volume} {58}},\ \bibinfo {pages} {096007} (\bibinfo
  {year} {1998})}\BibitemShut {NoStop}%
\bibitem [{\citenamefont {Scavenius}\ \emph {et~al.}(2001)\citenamefont
  {Scavenius}, \citenamefont {M\'ocsy}, \citenamefont {Mishustin},\ and\
  \citenamefont {Rischke}}]{ML10_16_e}%
  \BibitemOpen
  \bibfield  {author} {\bibinfo {author} {\bibfnamefont {O.}~\bibnamefont
  {Scavenius}}, \bibinfo {author} {\bibfnamefont {A.}~\bibnamefont {M\'ocsy}},
  \bibinfo {author} {\bibfnamefont {I.~N.}\ \bibnamefont {Mishustin}}, \ and\
  \bibinfo {author} {\bibfnamefont {D.~H.}\ \bibnamefont {Rischke}},\
  }\href@noop {} {\bibfield  {journal} {\bibinfo  {journal} {Phys. Rev. C}\
  }\textbf {\bibinfo {volume} {64}},\ \bibinfo {pages} {045202} (\bibinfo
  {year} {2001})}\BibitemShut {NoStop}%
\bibitem [{\citenamefont {Antoniou}\ and\ \citenamefont
  {Kapoyannis}(2003)}]{ML10_16_f}%
  \BibitemOpen
  \bibfield  {author} {\bibinfo {author} {\bibfnamefont {N.}~\bibnamefont
  {Antoniou}}\ and\ \bibinfo {author} {\bibfnamefont {A.}~\bibnamefont
  {Kapoyannis}},\ }\href@noop {} {\bibfield  {journal} {\bibinfo  {journal}
  {Phys. Lett. B}\ }\textbf {\bibinfo {volume} {563}},\ \bibinfo {pages} {165 }
  (\bibinfo {year} {2003})}\BibitemShut {NoStop}%
\bibitem [{\citenamefont {Hatta}\ and\ \citenamefont
  {Ikeda}(2003)}]{ML10_16_g}%
  \BibitemOpen
  \bibfield  {author} {\bibinfo {author} {\bibfnamefont {Y.}~\bibnamefont
  {Hatta}}\ and\ \bibinfo {author} {\bibfnamefont {T.}~\bibnamefont {Ikeda}},\
  }\href@noop {} {\bibfield  {journal} {\bibinfo  {journal} {Phys. Rev. D}\
  }\textbf {\bibinfo {volume} {67}},\ \bibinfo {pages} {014028} (\bibinfo
  {year} {2003})}\BibitemShut {NoStop}%
\bibitem [{\citenamefont {Stephanov}(2005)}]{ML17}%
  \BibitemOpen
  \bibfield  {author} {\bibinfo {author} {\bibfnamefont {M.~A.}\ \bibnamefont
  {Stephanov}},\ }\href@noop {} {\bibfield  {journal} {\bibinfo  {journal}
  {International Journal of Modern Physics A}\ }\textbf {\bibinfo {volume}
  {20}},\ \bibinfo {pages} {4387} (\bibinfo {year} {2005})}\BibitemShut
  {NoStop}%
\bibitem [{\citenamefont {Pratt}(1986)}]{RG96_a}%
  \BibitemOpen
  \bibfield  {author} {\bibinfo {author} {\bibfnamefont {S.}~\bibnamefont
  {Pratt}},\ }\href@noop {} {\bibfield  {journal} {\bibinfo  {journal} {Phys.
  Rev. D}\ }\textbf {\bibinfo {volume} {33}},\ \bibinfo {pages} {1314}
  (\bibinfo {year} {1986})}\BibitemShut {NoStop}%
\bibitem [{\citenamefont {Bertsch}\ \emph {et~al.}(1988)\citenamefont
  {Bertsch}, \citenamefont {Gong},\ and\ \citenamefont {Tohyama}}]{RG96_b}%
  \BibitemOpen
  \bibfield  {author} {\bibinfo {author} {\bibfnamefont {G.}~\bibnamefont
  {Bertsch}}, \bibinfo {author} {\bibfnamefont {M.}~\bibnamefont {Gong}}, \
  and\ \bibinfo {author} {\bibfnamefont {M.}~\bibnamefont {Tohyama}},\
  }\href@noop {} {\bibfield  {journal} {\bibinfo  {journal} {Phys. Rev. C}\
  }\textbf {\bibinfo {volume} {37}},\ \bibinfo {pages} {1896 } (\bibinfo {year}
  {1988})}\BibitemShut {NoStop}%
\bibitem [{\citenamefont {Rischke}\ and\ \citenamefont
  {Gyulassy}(1996)}]{RG96_c}%
  \BibitemOpen
  \bibfield  {author} {\bibinfo {author} {\bibfnamefont {D.~H.}\ \bibnamefont
  {Rischke}}\ and\ \bibinfo {author} {\bibfnamefont {M.}~\bibnamefont
  {Gyulassy}},\ }\href@noop {} {\bibfield  {journal} {\bibinfo  {journal}
  {Nucl. Phys. A}\ }\textbf {\bibinfo {volume} {608}},\ \bibinfo {pages} {479}
  (\bibinfo {year} {1996})}\BibitemShut {NoStop}%
\bibitem [{\citenamefont {Adare}\ \emph {et~al.}()\citenamefont {Adare} \emph
  {et~al.}}]{PHENIX_HBT}%
  \BibitemOpen
  \bibfield  {author} {\bibinfo {author} {\bibfnamefont {A.}~\bibnamefont
  {Adare}} \emph {et~al.},\ }\href@noop {} {\ }\Eprint
  {http://arxiv.org/abs/1410.2559} {arXiv:1410.2559 [nucl-ex]} \BibitemShut
  {NoStop}%
\bibitem [{\citenamefont {Karpenko}\ \emph {et~al.}(2015)\citenamefont
  {Karpenko}, \citenamefont {Huovinen}, \citenamefont {Petersen},\ and\
  \citenamefont {Bleicher}}]{vHLLE_UrQMD_Yura1}%
  \BibitemOpen
  \bibfield  {author} {\bibinfo {author} {\bibfnamefont {I.~A.}\ \bibnamefont
  {Karpenko}}, \bibinfo {author} {\bibfnamefont {P.}~\bibnamefont {Huovinen}},
  \bibinfo {author} {\bibfnamefont {H.}~\bibnamefont {Petersen}}, \ and\
  \bibinfo {author} {\bibfnamefont {M.}~\bibnamefont {Bleicher}},\ }\href@noop
  {} {\bibfield  {journal} {\bibinfo  {journal} {Phys. Rev. C}\ }\textbf
  {\bibinfo {volume} {91}},\ \bibinfo {pages} {064901} (\bibinfo {year}
  {2015})}\BibitemShut {NoStop}%
\bibitem [{\citenamefont {Bass}\ \emph {et~al.}(1998)\citenamefont {Bass} \emph
  {et~al.}}]{UrQMD_a}%
  \BibitemOpen
  \bibfield  {author} {\bibinfo {author} {\bibfnamefont {S.~A.}\ \bibnamefont
  {Bass}} \emph {et~al.},\ }\href@noop {} {\bibfield  {journal} {\bibinfo
  {journal} {Prog. Part. Nucl. Phys.}\ }\textbf {\bibinfo {volume} {41}},\
  \bibinfo {pages} {255} (\bibinfo {year} {1998})}\BibitemShut {NoStop}%
\bibitem [{\citenamefont {Bleicher}\ \emph {et~al.}(1999)\citenamefont
  {Bleicher} \emph {et~al.}}]{UrQMD_b}%
  \BibitemOpen
  \bibfield  {author} {\bibinfo {author} {\bibfnamefont {M.}~\bibnamefont
  {Bleicher}} \emph {et~al.},\ }\href@noop {} {\bibfield  {journal} {\bibinfo
  {journal} {J. Phys. G}\ }\textbf {\bibinfo {volume} {25}},\ \bibinfo {pages}
  {1859} (\bibinfo {year} {1999})}\BibitemShut {NoStop}%
\bibitem [{\citenamefont {Karpenko}\ \emph {et~al.}(2014)\citenamefont
  {Karpenko}, \citenamefont {Huovinen},\ and\ \citenamefont
  {Bleicher}}]{vHLLE}%
  \BibitemOpen
  \bibfield  {author} {\bibinfo {author} {\bibfnamefont {I.}~\bibnamefont
  {Karpenko}}, \bibinfo {author} {\bibfnamefont {P.}~\bibnamefont {Huovinen}},
  \ and\ \bibinfo {author} {\bibfnamefont {M.}~\bibnamefont {Bleicher}},\
  }\href@noop {} {\bibfield  {journal} {\bibinfo  {journal} {Comput. Phys.
  Commun.}\ }\textbf {\bibinfo {volume} {185}},\ \bibinfo {pages} {3016}
  (\bibinfo {year} {2014})}\BibitemShut {NoStop}%
\bibitem [{\citenamefont {Ivanov}\ \emph {et~al.}(2006)\citenamefont {Ivanov},
  \citenamefont {Russkikh},\ and\ \citenamefont {Toneev}}]{Ivanov}%
  \BibitemOpen
  \bibfield  {author} {\bibinfo {author} {\bibfnamefont {{\relax Yu}.~B.}\
  \bibnamefont {Ivanov}}, \bibinfo {author} {\bibfnamefont {V.~N.}\
  \bibnamefont {Russkikh}}, \ and\ \bibinfo {author} {\bibfnamefont {V.~D.}\
  \bibnamefont {Toneev}},\ }\href@noop {} {\bibfield  {journal} {\bibinfo
  {journal} {Phys. Rev. C}\ }\textbf {\bibinfo {volume} {73}},\ \bibinfo
  {pages} {044904} (\bibinfo {year} {2006})}\BibitemShut {NoStop}%
\bibitem [{\citenamefont {Cooper}\ and\ \citenamefont
  {Frye}(1974)}]{CooperFrye}%
  \BibitemOpen
  \bibfield  {author} {\bibinfo {author} {\bibfnamefont {F.}~\bibnamefont
  {Cooper}}\ and\ \bibinfo {author} {\bibfnamefont {G.}~\bibnamefont {Frye}},\
  }\href@noop {} {\bibfield  {journal} {\bibinfo  {journal} {Phys. Rev. D}\
  }\textbf {\bibinfo {volume} {10}},\ \bibinfo {pages} {186} (\bibinfo {year}
  {1974})}\BibitemShut {NoStop}%
\bibitem [{\citenamefont {Steinheimer}\ \emph {et~al.}(2011)\citenamefont
  {Steinheimer}, \citenamefont {Schramm},\ and\ \citenamefont
  {St{\"o}cker}}]{SchrammSteinheimer}%
  \BibitemOpen
  \bibfield  {author} {\bibinfo {author} {\bibfnamefont {J.}~\bibnamefont
  {Steinheimer}}, \bibinfo {author} {\bibfnamefont {S.}~\bibnamefont
  {Schramm}}, \ and\ \bibinfo {author} {\bibfnamefont {H.}~\bibnamefont
  {St{\"o}cker}},\ }\href@noop {} {\bibfield  {journal} {\bibinfo  {journal}
  {J. Phys. G}\ }\textbf {\bibinfo {volume} {38}},\ \bibinfo {pages} {035001}
  (\bibinfo {year} {2011})}\BibitemShut {NoStop}%
\bibitem [{\citenamefont {Kolb}\ \emph {et~al.}(2000)\citenamefont {Kolb},
  \citenamefont {Sollfrank},\ and\ \citenamefont {Heinz}}]{KolbSollfrankHeinz}%
  \BibitemOpen
  \bibfield  {author} {\bibinfo {author} {\bibfnamefont {P.~F.}\ \bibnamefont
  {Kolb}}, \bibinfo {author} {\bibfnamefont {J.}~\bibnamefont {Sollfrank}}, \
  and\ \bibinfo {author} {\bibfnamefont {U.~W.}\ \bibnamefont {Heinz}},\
  }\href@noop {} {\bibfield  {journal} {\bibinfo  {journal} {Phys. Rev. C}\
  }\textbf {\bibinfo {volume} {62}},\ \bibinfo {pages} {054909} (\bibinfo
  {year} {2000})}\BibitemShut {NoStop}%
\bibitem [{\citenamefont {Huovinen}\ and\ \citenamefont
  {Petersen}(2012)}]{Huovinen:2012is}%
  \BibitemOpen
  \bibfield  {author} {\bibinfo {author} {\bibfnamefont {P.}~\bibnamefont
  {Huovinen}}\ and\ \bibinfo {author} {\bibfnamefont {H.}~\bibnamefont
  {Petersen}},\ }\href@noop {} {\bibfield  {journal} {\bibinfo  {journal} {Eur.
  Phys. J. A}\ }\textbf {\bibinfo {volume} {48}},\ \bibinfo {pages} {171}
  (\bibinfo {year} {2012})}\BibitemShut {NoStop}%
\bibitem [{\citenamefont {Goldhaber}\ \emph {et~al.}(1960)\citenamefont
  {Goldhaber}, \citenamefont {Goldhaber}, \citenamefont {Lee},\ and\
  \citenamefont {Pais}}]{GGLP}%
  \BibitemOpen
  \bibfield  {author} {\bibinfo {author} {\bibfnamefont {G.}~\bibnamefont
  {Goldhaber}}, \bibinfo {author} {\bibfnamefont {S.}~\bibnamefont
  {Goldhaber}}, \bibinfo {author} {\bibfnamefont {W.}~\bibnamefont {Lee}}, \
  and\ \bibinfo {author} {\bibfnamefont {A.}~\bibnamefont {Pais}},\ }\href@noop
  {} {\bibfield  {journal} {\bibinfo  {journal} {Phys. Rev.}\ }\textbf
  {\bibinfo {volume} {120}},\ \bibinfo {pages} {300} (\bibinfo {year}
  {1960})}\BibitemShut {NoStop}%
\bibitem [{\citenamefont {Podgoretsky}(1989{\natexlab{a}})}]{pod89_a}%
  \BibitemOpen
  \bibfield  {author} {\bibinfo {author} {\bibfnamefont {M.}~\bibnamefont
  {Podgoretsky}},\ }\href@noop {} {\bibfield  {journal} {\bibinfo  {journal}
  {Fiz.Elem.Chast.Atom.Yadra}\ }\textbf {\bibinfo {volume} {20}},\ \bibinfo
  {pages} {628 } (\bibinfo {year} {1989}{\natexlab{a}})}\BibitemShut {NoStop}%
\bibitem [{\citenamefont {Podgoretsky}(1989{\natexlab{b}})}]{pod89_b}%
  \BibitemOpen
  \bibfield  {author} {\bibinfo {author} {\bibfnamefont {M.}~\bibnamefont
  {Podgoretsky}},\ }\href@noop {} {\bibfield  {journal} {\bibinfo  {journal}
  {Sov. J. Part. Nucl.}\ }\textbf {\bibinfo {volume} {20}},\ \bibinfo {pages}
  {266} (\bibinfo {year} {1989}{\natexlab{b}})}\BibitemShut {NoStop}%
\bibitem [{\citenamefont {Lednicky}(2004)}]{led04}%
  \BibitemOpen
  \bibfield  {author} {\bibinfo {author} {\bibfnamefont {R.}~\bibnamefont
  {Lednicky}},\ }\href@noop {} {\bibfield  {journal} {\bibinfo  {journal}
  {Phys. Atom. Nucl.}\ }\textbf {\bibinfo {volume} {67}},\ \bibinfo {pages}
  {72} (\bibinfo {year} {2004})}\BibitemShut {NoStop}%
\bibitem [{\citenamefont {Pratt}(1984)}]{lis05_a}%
  \BibitemOpen
  \bibfield  {author} {\bibinfo {author} {\bibfnamefont {S.}~\bibnamefont
  {Pratt}},\ }\href@noop {} {\bibfield  {journal} {\bibinfo  {journal} {Phys.
  Rev. Lett.}\ }\textbf {\bibinfo {volume} {53}},\ \bibinfo {pages} {1219}
  (\bibinfo {year} {1984})}\BibitemShut {NoStop}%
\bibitem [{\citenamefont {Lisa}\ \emph {et~al.}(2005)\citenamefont {Lisa},
  \citenamefont {Pratt}, \citenamefont {Soltz},\ and\ \citenamefont
  {Wiedemann}}]{lis05_b}%
  \BibitemOpen
  \bibfield  {author} {\bibinfo {author} {\bibfnamefont {M.~A.}\ \bibnamefont
  {Lisa}}, \bibinfo {author} {\bibfnamefont {S.}~\bibnamefont {Pratt}},
  \bibinfo {author} {\bibfnamefont {R.}~\bibnamefont {Soltz}}, \ and\ \bibinfo
  {author} {\bibfnamefont {U.}~\bibnamefont {Wiedemann}},\ }\href@noop {}
  {\bibfield  {journal} {\bibinfo  {journal} {Ann. Rev. Nucl. Part. Sci.}\
  }\textbf {\bibinfo {volume} {55}},\ \bibinfo {pages} {357} (\bibinfo {year}
  {2005})}\BibitemShut {NoStop}%
\bibitem [{\citenamefont {Lednick{\'y}}\ and\ \citenamefont
  {Lyuboshits}(1982)}]{Lednicky:1981su}%
  \BibitemOpen
  \bibfield  {author} {\bibinfo {author} {\bibfnamefont {R.}~\bibnamefont
  {Lednick{\'y}}}\ and\ \bibinfo {author} {\bibfnamefont {V.~L.}\ \bibnamefont
  {Lyuboshits}},\ }\href@noop {} {\bibfield  {journal} {\bibinfo  {journal}
  {Sov. J. Nucl. Phys.}\ }\textbf {\bibinfo {volume} {35}},\ \bibinfo {pages}
  {770} (\bibinfo {year} {1982})}\BibitemShut {NoStop}%
\bibitem [{\citenamefont {Lednick{\'y}}(2009)}]{Lednicky:2005tb}%
  \BibitemOpen
  \bibfield  {author} {\bibinfo {author} {\bibfnamefont {R.}~\bibnamefont
  {Lednick{\'y}}},\ }\href@noop {} {\bibfield  {journal} {\bibinfo  {journal}
  {Phys. Part. Nucl.}\ }\textbf {\bibinfo {volume} {40}},\ \bibinfo {pages}
  {307} (\bibinfo {year} {2009})}\BibitemShut {NoStop}%
\bibitem [{\citenamefont {Adamczyk}\ \emph {et~al.}(2013)\citenamefont
  {Adamczyk} \emph {et~al.}}]{SF}%
  \BibitemOpen
  \bibfield  {author} {\bibinfo {author} {\bibfnamefont {L.}~\bibnamefont
  {Adamczyk}} \emph {et~al.},\ }\href@noop {} {\bibfield  {journal} {\bibinfo
  {journal} {Phys. Rev. C}\ }\textbf {\bibinfo {volume} {88}},\ \bibinfo
  {pages} {034906} (\bibinfo {year} {2013})}\BibitemShut {NoStop}%
\bibitem [{\citenamefont {Karpenko}\ \emph {et~al.}(2016)\citenamefont
  {Karpenko}, \citenamefont {Bleicher}, \citenamefont {Huovinen},\ and\
  \citenamefont {Petersen}}]{Karpenko:2016red}%
  \BibitemOpen
  \bibfield  {author} {\bibinfo {author} {\bibfnamefont {I.}~\bibnamefont
  {Karpenko}}, \bibinfo {author} {\bibfnamefont {M.}~\bibnamefont {Bleicher}},
  \bibinfo {author} {\bibfnamefont {P.}~\bibnamefont {Huovinen}}, \ and\
  \bibinfo {author} {\bibfnamefont {H.}~\bibnamefont {Petersen}},\ }\href@noop
  {} {\bibfield  {journal} {\bibinfo  {journal} {Nucl. Phys. A}\ }\textbf
  {\bibinfo {volume} {956}},\ \bibinfo {pages} {834} (\bibinfo {year}
  {2016})}\BibitemShut {NoStop}%
\bibitem [{\citenamefont {Adamczyk}\ \emph {et~al.}(2015)\citenamefont
  {Adamczyk} \emph {et~al.}}]{STAR_HBT}%
  \BibitemOpen
  \bibfield  {author} {\bibinfo {author} {\bibfnamefont {L.}~\bibnamefont
  {Adamczyk}} \emph {et~al.},\ }\href@noop {} {\bibfield  {journal} {\bibinfo
  {journal} {Phys. Rev. C}\ }\textbf {\bibinfo {volume} {92}},\ \bibinfo
  {pages} {014904} (\bibinfo {year} {2015})}\BibitemShut {NoStop}%
\bibitem [{\citenamefont {Alt}\ \emph {et~al.}(2010)\citenamefont {Alt} \emph
  {et~al.}}]{Alt:HumpFunction}%
  \BibitemOpen
  \bibfield  {author} {\bibinfo {author} {\bibfnamefont {C.}~\bibnamefont
  {Alt}} \emph {et~al.},\ }\href@noop {} {\bibfield  {journal} {\bibinfo
  {journal} {Phys. Lett.}\ }\textbf {\bibinfo {volume} {B685}},\ \bibinfo
  {pages} {41} (\bibinfo {year} {2010})}\BibitemShut {NoStop}%
\bibitem [{\citenamefont {Batyuk}\ \emph {et~al.}(2016)\citenamefont {Batyuk},
  \citenamefont {Blaschke}, \citenamefont {Bleicher}, \citenamefont {Ivanov},
  \citenamefont {Karpenko}, \citenamefont {Merts}, \citenamefont {Nahrgang},
  \citenamefont {Petersen},\ and\ \citenamefont
  {Rogachevsky}}]{Batyuk:THESEUS}%
  \BibitemOpen
  \bibfield  {author} {\bibinfo {author} {\bibfnamefont {P.}~\bibnamefont
  {Batyuk}}, \bibinfo {author} {\bibfnamefont {D.}~\bibnamefont {Blaschke}},
  \bibinfo {author} {\bibfnamefont {M.}~\bibnamefont {Bleicher}}, \bibinfo
  {author} {\bibfnamefont {{\relax Yu}.~B.}\ \bibnamefont {Ivanov}}, \bibinfo
  {author} {\bibfnamefont {I.}~\bibnamefont {Karpenko}}, \bibinfo {author}
  {\bibfnamefont {S.}~\bibnamefont {Merts}}, \bibinfo {author} {\bibfnamefont
  {M.}~\bibnamefont {Nahrgang}}, \bibinfo {author} {\bibfnamefont
  {H.}~\bibnamefont {Petersen}}, \ and\ \bibinfo {author} {\bibfnamefont
  {O.}~\bibnamefont {Rogachevsky}},\ }\href@noop {} {\bibfield  {journal}
  {\bibinfo  {journal} {Phys. Rev. C}\ }\textbf {\bibinfo {volume} {94}},\
  \bibinfo {pages} {044917} (\bibinfo {year} {2016})},\ \Eprint
  {http://arxiv.org/abs/1608.00965} {arXiv:1608.00965 [nucl-th]} \BibitemShut
  {NoStop}%
\end{thebibliography}%
\end{document}